\documentclass[eqsecnum,amsmath,showpacs,amsfonts,aps,prc,floatfix]{revtex4} 
\usepackage{graphicx}
\usepackage{bm}

\begin{document}

 \title{Viscous fluid dynamics in Au+Au collisions at RHIC}

\author{A. K. Chaudhuri}
\affiliation{Variable Energy Cyclotron Centre, 1/AF, Bidhan Nagar, 
Kolkata 700~064, India}

\begin{abstract}
  We have studied the space-time evolution of minimally viscous ($\frac{\eta}{s}$=0.08) QGP fluid, undergoing boost-invariant longitudinal motion and arbitrary transverse expansion.
Relaxation equations for the shear stress tensor components, 
derived from the phenomenological Israel-Stewart's theory of dissipative relativistic fluid, are   solved simultaneously with the energy-momentum conservation equations. 
Comparison of evolution of ideal and viscous fluid, both
initialized under the similar conditions, e.g. same equilibration time, energy density and velocity profile, indicate that in  viscous fluid, energy density or temperature of the fluid evolve slowly than in an ideal fluid. Transverse expansion is also more in viscous evolution. We have also studied particle production in viscous dynamics. Compared to ideal dynamics, in viscous dynamics, particle yield at high $p_T$ is increased. Elliptic flow on the other hand decreases.  
Minimally viscous QGP fluid,
initialized at entropy density $s_{ini}$=110 $fm^{-3}$ at the initial time $\tau_i$=0.6 fm, if freeze-out at temperature $T_F$=130 MeV, explains the centrality dependence of $p_T$ spectra of identified particles. Experimental $p_T$ spectra of $\pi^-$, $K^+$ and protons in 0-5\%, 5-10\%, 10-20\%, 20-30\%, 30-40\% and 40-50\% Au+Au collisions are well reproduced throughout the experimental $p_T$ range. This is in contrast to ideal dynamics, where, the spectra are reproduced only up to $p_T\approx$1.5 GeV.  Minimally viscous QGP fluid, also explain the elliptic flow in mid-central (10-20\%, 16-23\%, 20-30\%) collisions. The minimum bias elliptic flow is also explained. However, the model under-predict/over-predict  the elliptic flow in very central/peripheral collisions.  
\end{abstract}


\date{\today}  

\maketitle

\section{Introduction}
\label{sec1}
%
Recent Au+Au collisions at the Relativistic Heavy Ion Collider (RHIC), at  cm energy of $\sqrt{s_{NN}}$=200 GeV, give a strong indication that in central collisions, a hot dense matter is formed
\cite{BRAHMSwhitepaper,PHOBOSwhitepaper,PHENIXwhitepaper,STARwhitepaper} .
Whether the formed matter can be identified as   the much sought after Quark-Gluon Plasma (QGP) as predicted in Lattice QCD simulations \cite{lattice} is presently debatable. 
However, experimentally observed elliptic flow in non-central Au+Au collisions gave a strong indication that a
thermalized collective QCD matter is formed.  Additionally,
success of {\em ideal} fluid dynamics 
in explaining 
several experimental data e.g. transverse momentum spectra of identified particles, elliptic flow etc \cite{QGP3}, together with the string theory motivated lower limit of shear viscosity $\frac{\eta}{s} \geq 1/4\pi$ \cite{Policastro:2001yc,Policastro:2002se} leading to a paradigm that in Au+Au collisions, a nearly perfect fluid is created.  

However, the paradigm of "perfect fluid" produced in Au+Au collisions at RHIC need to be clarified. It so happens that
the ideal fluid dynamic models do have their shortcomings \cite{Heinz:2004ar}. For example, experimentally, elliptic flow tends to saturate at large transverse momentum. The ideal fluid dynamics on the other hand predicts a continually increasing elliptic flow. The transverse momentum spectra 
of identified particles also starts to deviate from ideal fluid dynamics prediction beyond $p_T\approx$ 1.5 GeV. Experimentally determined  HBT radii are not reproduced in the ideal fluid dynamic models, the famous "HBT puzzle" \cite{Heinz:2002un}.  
Ideal fluid dynamics also works best in central collisions and gets poorer in more peripheral collisions.  

The shortcomings of ideal fluid dynamics possibly indicate
greater importance of dissipative effects  in the $p_T$ ranges greater than 1.5 GeV or in more peripheral collisions. Indeed, 
ideal fluid is a concept, which is never realized in nature. As suggested
in string theory motivated models,  QGP viscosity could be small,   $\eta/s \geq 1/4\pi$, nevertheless it is non-zero. It is important
to study the effect of viscosity, even if small, on space-time evolution of QGP fluid and quantify its effect. This requires a numerical implementation of relativistic dissipative fluid dynamics. Furthermore, 
if QGP fluid is formed in heavy ion collisions, it has to be characterized by measuring its transport coefficients, e.g. heat conductivity, bulk and shear viscosity. 
Theoretically, it is possible to obtain those transport coefficients in a kinetic theory model. However,
in the present status of theory,  the goal can not be achieved immediately, even more so for a strongly interacting QGP (sQGP).
Alternatively, one can use the
experimental data to obtain a "phenomenological" limit to the transport coefficients of sQGP. 
It will also require a numerical implementation of relativistic dissipative fluid dynamics. There is another incentive to study 
dissipative   hydrodynamics. Ideal hydrodynamics depends on the
assumption of local equilibrium.  
Dissipative hydrodynamics on the other hand do not depend on the assumption of local equilibrium. As will be discussed later, the fluid should not be far from an equilibrium state. The range of validity of dissipative hydrodynamics thus increases. Indeed, we can explore early times of fluid evolution better in a dissipative hydrodynamics. 

Theory of dissipative relativistic 
fluid has been formulated quite early.  The original dissipative relativistic fluid equations were given by Eckart \cite{Eckart}
and Landau and Lifshitz \cite{LL63}. They are called 1st order theories. Formally, relativistic dissipative hydrodynamics are obtained from an expansion of entropy 4-current, in terms of dissipative fluxes. In 1st order theories, entropy 4-current contains terms linear in dissipative quantities.   
1st order theory of dissipative hydrodynamics 
suffers from the problem of causality violation. Signal can travel faster than light.  Causality violation is unwarranted in any theory,
even more in a relativistic theory. 
 The problem of causality violation is removed in the Israel-Stewart's 2nd order theory of dissipative fluid \cite{IS79}. In 2nd order theory, 
expansion of entropy 4-current contains terms 2nd order in dissipative fluxes. However, these leads to complications that
dissipative fluxes are no longer function of the state variables only.
They become dynamic.
The space of  thermodynamic variables has to be extended to include the dissipative fluxes (e.g. heat conductivity, bulk and shear viscosity).

Even though 2nd order theory is formulated some 30 years back, significant progress towards its numerical implementation 
has only been made very recently 
\cite{Teaney:2004qa,Teaney:2003kp,Muronga:2001zk,Muronga:2006zx,Muronga:2006zw,Muronga:2004sf,Song:2007fn,Song:2007ux,Heinz:2005bw,Chaudhuri:2005ea,Chaudhuri:2006jd,Chaudhuri:2007yn,Chaudhuri:2007yk,Chaudhuri:2007zm,Chaudhuri:2007qp,Romatschke:2007mq,Baier:2006gy,Baier:2006um,Koide:2007kw,Denicol:2007zz}.
At the Cyclotron Centre, 
Kolkata, we have developed a code  
"AZHYDRO-KOLKATA"  to simulate the  
hydrodynamic evolution of QGP fluid including the effect of  dissipation due to shear viscosity only. The code can simulate both
the 1st order and the 2nd order dissipative hydrodynamics.
Some results of AZHYDRO-KOLKATA  for first order dissipative hydrodynamics have been published earlier \cite{Chaudhuri:2006jd,Chaudhuri:2007yn,Chaudhuri:2007yk,Chaudhuri:2007zm}.
In an earlier paper \cite{Chaudhuri:2007zm}, we have studied the 2nd order dissipative hydrodynamics . The study was limited to the QGP phase only.  More recently, the code was extended to include the phase transition \cite{Chaudhuri:2007qp}. 
In the present paper, we study, in detail, 2nd order dissipative  hydrodynamics evolution of QGP fluid including  phase transition. We will also compare viscous hydrodynamics predictions with experimental data. It will be shown that the
experimental $p_T$ spectra of identified particles and elliptic flow are much better explained in
2nd order dissipative hydrodynamics  than in ideal hydrodynamics.
 
The paper is organized as follows: In section~\ref{sec2} we shortly
review the Israel-Stewart's \cite{IS79} phenomenological theory of relativistic dissipative fluid dynamics. We will limit our study to boost-invariant motion and consider dissipation only due to the shear viscosity.
In section~\ref{sec3} we derive the relevant equations in 2+1 dimension. In section \ref{sec4}, we 
discuss the equation of state, viscosity coefficient and initial conditions used in the present study. Hydrodynamic evolution gives the temporal evolution of thermodynamics variables, e.g. energy density or temperature, fluid velocity components, the shear stress tensor
components. The information needs to be converted to particle spectra to compare against experiments. In \ref{sec5},
we present the relevant equations for computing invariant
particle distribution in viscous dynamics.  We test the numerical accuracy of the code "AZHYDRO-KOLKATA" in section \ref{sec6}. Evolution of ideal and minimally viscous fluid is compared in section \ref{sec7}. In section \ref{sec8} $p_T$ spectra and elliptic flow in viscous dynamics are compared with the experimental data.
Fluid evolution and subsequent particle production depend on
the initial shear stress tensor and also on the relaxation time.
Some preliminary study is described in   section~\ref{sec9a}.
The concluding section~\ref{sec9} summarizes our results.

\section{dissipative fluid dynamics}
\label{sec2}
%

In this section, I briefly discuss the Israel-Stewarts phenomenological theory of
dissipative hydrodynamics. More detailed exposition can be found in \cite{IS79}.  

A simple fluid, in an arbitrary state, is fully specified by primary variables: particle current ($N^\mu$), energy-momentum tensor ($T^{\mu\nu}$) and
entropy current ($S^\mu$) and a number of additional (unknown) variables. Primary variables satisfies the conservation
laws;

\begin{eqnarray} 
\partial_\mu N^\mu =&&0,\label{eq1}\\
\partial_\mu T^{\mu\nu}=&&0, \label{eq2}
\end{eqnarray} 

\noindent and the 2nd law of thermodynamics,

\begin{equation}
\partial_\mu S^\mu  \geq 0. \label{eq3}
\end{equation}

In relativistic fluid dynamics, one defines a time-like hydrodynamic 4-velocity,
$u^\mu$ (normalized as $u^2=1$). One also define 
a projector, 
$\Delta^{\mu\nu}=g^{\mu\nu}-u^\mu u^\nu$,
orthogonal to the 4-velocity ($\Delta^{\mu\nu}u_\nu=0$).
In equilibrium, an unique
4-velocity ($u^\mu$) exists such that the particle density ($n$), energy density ($\varepsilon$) and the entropy density ($s$) can be obtained from,

\begin{eqnarray} \label{eq4}
 N^\mu_{eq}=&& n u^\mu  \\
\label{eq5}
T^{\mu\nu}_{eq}=&&\varepsilon u^\mu u^\nu -p \Delta^{\mu\nu}\\
\label{eq6}
S^\mu_{eq}=&&s u^\mu 
\end{eqnarray}

An equilibrium state is assumed to be fully specified by 5-parameters,
$(n,\varepsilon,u^\mu)$ or equivalently by the thermal potential,
$\alpha=\mu/T$ ($\mu$ being the chemical potential) and inverse 4-temperature, $\beta^\mu=u^\mu/T$. Given a equation of state, $s=s(\varepsilon,n)$, pressure $p$ can be obtained from the generalized thermodynamic relation,

\begin{equation} \label{eq7}
S^\mu_{eq}=p\beta^\mu-\alpha N^\mu_{eq} +\beta_\lambda T^{\lambda\mu}_{eq}
\end{equation} 

Using the Gibbs-Duhem relation, 
$d(p\beta^\mu)=N^\mu_{eq} d\alpha -T^{\lambda\mu}_{eq}d\beta_\lambda$, following relations can be established on the equilibrium hyper-surface $\Sigma_{eq}(\alpha,\beta^\mu)$,


\begin{equation} \label{eq8}
dS^\mu_{eq}=-\alpha dN^\mu_{eq}+\beta_\lambda dT^{\lambda\mu}_{eq}
\end{equation}

In a non-equilibrium system, no 4-velocity can be found such that Eqs.\ref{eq4},\ref{eq5},\ref{eq6} remain valid. Tensor decomposition leads to additional terms,  

\begin{eqnarray}\label{eq9}
N^\mu=&&N^\mu_{eq}+\delta N^\mu=nu^\mu + V^\mu\\
\label{eq10}
T^{\mu\nu} =&&T^{\mu\nu}_{eq}+\delta T^{\mu\nu} \nonumber \\
=&&
[\varepsilon u^\mu u^\nu-p\Delta^{\mu\nu}]+\Pi\Delta^{\mu\nu} + \pi^{\mu\nu} \nonumber\\
&&+(W^\mu u^\nu + W^\nu u^\mu)\\
\label{eq11}
S^\mu=&&S^\mu_{eq}+\delta S^\mu=su^\mu + \Phi^\mu
\end{eqnarray} 

The new terms describe a net flow of charge $V^\mu=\Delta^{\mu\nu} N_\nu$, heat flow, $W^\mu=(\varepsilon+p)/n V^\mu +q^\mu$ (where $q^\mu$ is the heat flow vector), and entropy flow $\Phi^\mu$. 
$\Pi=-\frac{1}{3}\Delta_{\mu\nu}T^{\mu\nu}-p$ is the bulk viscous pressure
and $\pi^{\mu\nu}= [\frac{1}{2}(\Delta^{\mu\sigma}\Delta^{\nu\tau}+ 
\Delta^{\nu\sigma}\Delta^{\mu\tau}-\frac{1}{3}
\Delta^{\mu\nu}\Delta^{\sigma\tau}]T_{\sigma\tau}$ is the shear stress tensor.
Hydrodynamic 4-velocity can be chosen
to eliminate either $V^\mu$ (the Eckart frame, $u^\mu$ is parallel 
to particle flow) or the heat flow $q^\mu$ (the Landau frame, $u^\mu$ is
parallel to energy flow). In relativistic heavy ion collisions,
central rapidity region is nearly baryon free and Landau's frame is more appropriate than the Eckart's frame. Dissipative flows are transverse to $u^\mu$ and additionally, shear stress tensor is traceless. Thus a non-equilibrium state  require 1+3+5=9 additional quantities, the dissipative 
flows $\Pi$, $q^\mu$ (or $V^\mu$) and $\pi^{\mu\nu}$.  
In kinetic theory, $N^\mu$ and $T^{\mu\nu}$ are the 1st and 2nd moment of the distribution function. Unless the function is known a-priori, two moments do not furnish enough information to enumerate the microscopic states required to determine $S^\mu$, and
in an arbitrary non-equilibrium state, no relation exists between,
$N^\nu$, $T^{\mu\nu}$ and $S^\mu$. 
{\em Only in a state, close to 
an equilibrium one, such a relation can be established}. 
Assuming that the equilibrium relation Eq.\ref{eq8} remains valid
in a "near equilibrium state" also, the entropy current can be generalized as,

\begin{equation} \label{eq12}
S^\mu=S^\mu_{eq}+dS^\mu
=p\beta^\mu-\alpha N^\mu +\beta_\lambda T^{\lambda\mu} + Q^\mu
\end{equation}

\noindent where $Q^\mu$ is an undetermined quantity in 2nd order in deviations, $\delta N^\mu=N^\mu-N^\mu_{eq}$ and $\delta T^{\mu\nu}=T^{\mu\nu}-T^{\mu\nu}_{eq}$.  Detail form of $Q^\mu$ is constrained by the 2nd law $\partial_\mu S^\mu \geq 0$.
With the help of conservation laws and Gibbs-Duhem relation,
entropy production rate can be written as,

\begin{eqnarray} \label{eq13}
\partial_\mu S^\mu=-\delta N^\mu \partial_\mu \alpha
+\delta T^{\mu\nu} \partial_\mu \beta_\nu + \partial_\mu Q^\mu
\end{eqnarray}

Choice of $Q^\mu$ leads to 1st order or 2nd order theories of dissipative hydrodynamics.
In 1st order theories the simplest choice is made, $Q^\mu=0$, entropy current contains terms up to 1st order in deviations,
$\delta N^\mu$ and $\delta T^{\mu\nu}$. Entropy production rate can be written as,

\begin{equation}\label{eq14}
T\partial_\mu S^\mu
=\Pi X -q^\mu X_\mu + \pi^{\mu\nu} X_{\mu\nu}  
\end{equation}

\noindent where, $X=-\nabla.u$; $X^\mu=\frac{\nabla^\mu}{T}-u^\nu \partial_\nu u^\mu$ 
and 
$X^{\mu\nu}=\nabla^{<\mu} u^{\nu>}$.
   
The 2nd law, $\partial_\mu S^\mu \geq 0$ can be satisfied by postulating a linear relation between the dissipative flows and thermodynamic forces,
 
\begin{eqnarray}
\label{eq15}
\Pi=&&-\zeta \theta,\\
\label{eq16}
q^\mu=&&-\lambda \frac{nT^2}{\varepsilon+p}\nabla^\mu(\mu/T),\\ 
\label{17}
\pi^{\mu\nu}=&&2\eta \nabla^{<\mu}u^{\nu>}
\end{eqnarray}

\noindent where $\zeta$, $\lambda$ and $\eta$ are the positive transport coefficients, bulk viscosity, heat conductivity and shear viscosity respectively. 

In 1st order theories, causality is violated. If, in a given fluid cell, at a certain time, thermodynamic forces vanish, corresponding dissipative fluxes also vanish instantly. Violation of causality is unwanted in any theory, even more so in relativistic theory. Causality violation of dissipative hydrodynamics is corrected in   2nd order theories \cite{IS79}. In 2nd order theories, entropy current contain terms up to 2nd order in the deviations, $Q^\mu \neq 0$. 
The most general $Q^\mu$ containing terms up to 2nd order in deviations can be written as,

\begin{equation} \label{eq18}
Q^\mu=-(\beta_0\Pi^2-\beta_1 q^\nu q_\nu + \beta_2\pi_{\nu\lambda}\pi^{\nu\lambda})
\frac{u^\mu}{2T} -\frac{\alpha_0\Pi q^\mu}{T} +\frac{\alpha_1 \pi^{\mu\nu}q_\nu}{T}
\end{equation}

As before, one can cast the entropy production rate ($T\partial_\mu S^\mu$) in the form of Eq.\ref{eq14}. 
Neglecting the terms involving dissipative flows with gradients of equilibrium thermodynamic quantities (both are assumed to be small) and demanding that a linear
relation exists between the dissipative flows and thermodynamic
forces,  following {\em relaxation} equations for the dissipative flows can be obtained,  

\begin{eqnarray} \label{eq19}
\Pi&=&-\zeta (\theta +\beta_0 D\Pi)\\
\label{eq20}
q^\mu&=&-\lambda \left[ \frac{nT^2}{\varepsilon+p}\nabla^\mu(\frac{\mu}{T}) 
-\beta_1 Dq^\mu \right]\\
\label{eq21}
\pi^{\mu\nu}&=&2\eta \left[\nabla^{<\mu}u^{\nu>} -\beta_2 D\pi^{\mu\nu} \right],
\end{eqnarray}

\noindent where  $D=u^\mu \partial_\mu$is the convective time derivative. Unlike in the 1st order theories, in 2nd order theories,
dynamical equations control the dissipative flows.  
Even if thermodynamic forces vanish, dissipative flows do not vanish instantly. It is important to mention that the parameters,
$\alpha$ and $\beta_\lambda$ are not connected to the actual state
($N^\mu,T^{\mu\nu}$). The pressure $p$ in Eq.\ref{eq12} is also not
the "actual" thermodynamics pressure, i.e. not the work done in
an isentropic expansion. Chemical potential $\alpha$ and 4-inverse temperature $\beta_\lambda$ has meaning only for the equilibrium state. Their meaning need not be extended to non-equilibrium states also. However, it is possible to fit a fictitious "local equilibrium" state, point by point, such that pressure $p$ in
Eq.\ref{eq12} can be identified with the thermodynamic pressure,  
at least up to  1st order. The conditions of fit fixes the  
underlying non-equilibrium phase-space distribution.

It  may be mentioned here that relaxation equations for the dissipative fluxes can also be derived in kinetic theory \cite{deGroot,Baier:2006um}. Relaxation equation from kinetic theory can contain additional terms, which are missed in the
Israel-Stewart's theory. In \cite{Song:2007fn,Song:2007ux,Romatschke:2007mq,Baier:2006gy}, authors have used 
the following relaxation equation, derived from the kinetic theory,

 \begin{equation} \label{eqkt}
\pi^{\mu\nu}=2\eta \left[\nabla^{<\mu}u^{\nu>} -\tau_\pi D\pi^{\mu\nu} \right]
- \tau_\pi [u^\mu\pi^{\nu\lambda}+u^\nu\pi^{\mu\lambda}]Du_\lambda
\end{equation}

\noindent  which contain an additional term $R=-[u^\mu\pi^{\nu\lambda}+u^\nu\pi^{\mu\lambda}]Du_\lambda$ vis-a-vis Israel-Stewart's relaxation equation. The additional term  ensures  
that throughout the evolution shear stress tensor remains traceless and transverse to fluid velocity. Israel-Stewart's theory is based on gradient expansion of entropy, gradients of
equilibrium thermodynamic quantities ($\varepsilon$, $u$) are assumed to be small. The term R does not contribute to entropy production and is missed in Eq.\ref{eq21}.   In the present work, we use Israel-Stewart's relaxation equation (neglecting the term R) to compute evolution of QGP fluid.
As assumed in Israel-Stewart's theory, if gradients of thermodynamical quantities are small   and since in principle, non-equilibrium effects are supposed to be small, both $Du_\mu$ and $\pi^{\mu\nu}$ small are small quantities and their product  $\pi^{\mu\nu}Du_\mu$ can contribute only in 2nd order. 
It will be shown later that  the contribution of the term R in hydrodynamic evolution of minimally viscous fluid is negligible. The evolution of energy density of the fluid is hardly changed whether or not the term R is included in the relaxation equation. However, for more viscous fluid,
relaxation equation Eq.\ref{eqkt} will be more appropriate than Israel-Stewart's 
equation \ref{eq21}.

\section{(2+1)-dimensional viscous hydrodynamics with
longitudinal boost invariance}
\label{sec3}
Complete dissipative hydrodynamics is a numerically challenging problem. It requires simultaneous solution of 14 partial differential equations (5 conservation equations and 9 relaxation equations for dissipative flows). We reduce the problem to solution of 6 partial differential equations 
(3 conservation equations and 3 relaxation equations).
In the following, we will study boost-invariant  evolution of baryon free QGP fluid, including the dissipative effect due to shear viscosity only. Shear viscosity is the most important dissipative effect.  
For example, in  a baryon free QGP, heat conduction is zero and we can disregard Eq.\ref{eq20}. Bulk viscosity is also zero for the QGP fluid (point particles) and Eq.\ref{eq19}
can also be neglected. Shear pressure tensor has 5 independent 
components but the assumption of boost invariance reduces the
number of independent components to three. 
For a baryon free fluid, we can also disregard the conservation equation Eq.\ref{eq1}.  With the assumption of boost-invariance,
 energy-momentum conservation equation  $\partial_\mu T^{\mu \eta}=0$  become redundant.  

Heavy ion collisions
are best described in ($\tau,x,y,\eta$) coordinates, where $\tau=\sqrt{t^2-z^2}$ is the longitudinal proper time and  $\eta=\frac{1}{2} \ln \frac{t+z}{t-z}$ is the space-time rapidity.
$r_\perp=(x,y)$ are the usual cartisan coordinate in the plane, 
transverse to the beam direction. 
With longitudinal boost-invariance   the 
energy-momentum conservation equations ${T^{mn}}_{;n}=0$ yield 

 \begin{widetext}
\begin{eqnarray}
\label{eqb1}
&& \partial_\tau \tilde{T}^{\tau\tau}  
 +\partial_x (\tilde{T}^{\tau \tau} \overline{v}_x  )+\partial_y  (\tilde{T}^{\tau \tau} \overline{v}_y)= 
  -\,(p+\tau^2 \pi^{\eta\eta})
\\
\label{eqb2}
&&\partial_\tau\tilde{T}^{\tau x} 
 +\partial_x (\tilde{T}^{\tau x}v_x)
 +\partial_y (\tilde{T}^{\tau x}v_y) 
 = -\partial_x(\tilde{p} + \tilde{\pi}^{xx}-\tilde{\pi}^{\tau x} v_x) - \partial_y(\tilde{\pi}^{xy}-\tilde{\pi}^{\tau x}v_y)
\\
\label{eqb3}
 &&\partial_\tau\tilde{T}^{\tau y} 
 +\partial_x (\tilde{T}^{\tau y}v_x)
 +\partial_y (\tilde{T}^{\tau y}v_y) 
 = -\partial_x(\tilde{\pi}^{xy}-\tilde{\pi}^{\tau y} v_x) - \partial_y(\tilde{p} + \tilde{\pi}^{yy}-\tilde{\pi}^{\tau y}v_y)
\end{eqnarray}
\end{widetext}

\noindent where $\tilde{A}^{mn}\equiv  \tau A^{mn}$,
$\tilde{p}\equiv \tau p$, and 
$\overline{v}_x \equiv T^{\tau x}/T^{\tau\tau}$,
$\overline{v}_y \equiv T^{\tau y}/T^{\tau\tau}$.

The components of the energy momentum tensor, including the shear pressure tensor are, 

\begin{eqnarray}
\label{eqb4}
T^{\tau\tau} =&& (\varepsilon+p)\gamma_\perp^2 - p + \pi^{\tau\tau}\\
\label{eqb5}
T^{\tau x} =&& (\varepsilon+p)\gamma_\perp^2 v_x + \pi^{\tau x}\\
\label{eqb6}
T^{\tau y} =&& (\varepsilon+p)\gamma_\perp^2 v_y + \pi^{\tau y}
\end{eqnarray}
 
We note that unlike in ideal fluid, in viscous fluid dynamics, conservation equations (see Eqs.\ref{eqb1}-\ref{eqb3}) contain additional pressure gradients due to shear viscosity.  Both $T^{\tau x}$ and $T^{\tau y}$ components of energy-momentum tensor now evolve under additional pressure gradients. The rightmost term of Eq.\ref{eqb1} also indicate that in viscous dynamics, longitudinal pressure  is 
effectively reduced (note that the $\pi^{\eta\eta}$ component is negative). Since pressure can not be negative, shear viscosity is limited by the condition, $p + \tau^2\pi^{\eta\eta} \geq 0$. The energy momentum conservation equations \ref{eqb1}-\ref{eqb3} contain seven shear stress tensor components,
$\pi^{xx}$, $\pi^{yy}$, $\pi^{xy}$, $\pi^{\eta\eta}$, $\pi^{\tau\tau}$, $\pi^{\tau x}$ and $\pi^{\tau y}$. In 2nd order theory, the shear stress tensor is dynamical and evolves with time and the energy momentum conservation equations are to be solved together with the relaxation equations for the shear stress tensor,  As mentioned earlier, with boost-invariance, only three stress tensor components are independent. We choose $\pi^{xx}$, $\pi^{yy}$ and $\pi^{xy}$ as the independent components. The dependent shear stress tensor components can be expressed in terms of the independent ones. The relaxation equations for the independent shear stress tensor components,
in $(\tau,x,y,\eta)$ co-ordinates  can be written as,

\begin{widetext}
\begin{eqnarray}
\label{eqc4a}
\partial_\tau \pi^{xx} +v_x \partial_x \pi^{xx}+v_y \partial_y \pi^{xx}
&=&
-\frac{1}{\tau_\pi \gamma} \left (\pi^{xx} - 2\eta \sigma^{xx}\right )\\
\label{eqc5}
\partial_\tau \pi^{yy} +v_x \partial_x \pi^{yy}+v_y \partial_y \pi^{yy}
&=&
-\frac{1}{\tau_\pi \gamma} \left (\pi^{yy} - 2\eta \sigma^{yy}\right )\\
\label{eqc6}
\partial_\tau \pi^{xy} +v_x \partial_x \pi^{xy}+v_y \partial_y \pi^{xy}
&=&
-\frac{1}{\tau_\pi \gamma} \left (\pi^{xy} - 2\eta \sigma^{xy}\right )
\end{eqnarray}
\end{widetext} 
 
\noindent 
where $\tau_\pi$ is the relaxation time, $\tau_\pi=2\eta \beta_2$ (see
Eq.\ref{eq21}). 
The viscous pressure tensor relaxes on a   time scale $\tau_\pi$
to $2\eta$ times the shear tensor $\sigma^{\mu\nu}=
\nabla^{\left\langle\mu\right.}u^{\left.\nu\right\rangle}$.
The $xx$, $yy$  and $xy$ components of the shear tensor $\sigma^{\mu\nu}$ can be written as
%
\begin{eqnarray}
\label{eqc7}
\sigma^{xx}
=&&-\partial_x u^x -u^x Du^x -\frac{1}{3} \Delta^{xx} \theta\\ 
\label{eqc8}
\sigma^{yy}
=&&-\partial_y u^y -u^y Du^y -\frac{1}{3} \Delta^{yy} \theta\\ 
\label{eqc9}
 \sigma^{xy}
=&&-\frac{1}{2}[\partial_x u^y - \partial_y u^x
-u^x Du^y - u^y Du^x] \nonumber \\
&& -\frac{1}{3} \Delta^{xy} \theta
\end{eqnarray}

Here $Du^\mu=u^\mu \partial_\mu$ is the convective time derivative and $\theta=\partial_\mu u^\mu$ is the expansion scalar.  

The dependent shear stress tensor components can easily be obtained from
the independent ones. Using the properties that (i) $\pi^{\mu\nu}$ is transverse to $u^\mu$ and (ii) $\pi^{\mu\nu}$ is traceless, $g^{\mu\nu}\pi_{\mu\nu}=0$,     the dependent shear stress tensor components can be obtained as,
  
\begin{eqnarray}
\pi^{\tau x}=&& v_x \pi^{xx} + v_y \pi^{xy} \label{eqc1} \\
\pi^{\tau y}=&& v_x \pi^{xy} + v_y \pi^{yy} \label{eqc2}\\
\pi^{\tau \tau}=&& v^2_x \pi^{xx} + v^2_y \pi^{yy}+2v_x v_y \pi^{xy} \label{eqc3}\\
\tau^2 \pi^{\eta\eta}=&&   
-(1-v^2_x)\pi^{xx} - (1-v^2_y)\pi^{yy} \nonumber\\
&&+2v_xv_y\pi^{xy} \label{eqc4}
\end{eqnarray}

In causal dissipative hydrodynamics, energy momentum conservation equations [Eqs.\ref{eqb1}-\ref{eqb3}] are solved simultaneously with the
relaxation equations [Eqs.\ref{eqc4a}-\ref{eqc6}].
In ideal hydrodynamics, the energy-momentum conservation equations are closed with an equation of state and given initial conditions, e.g. energy density ($\varepsilon$) and fluid velocity ($v_x$ and $v_y$) distributions at time $\tau_i$, the equation of motions can be integrated to obtain the variables at the next time step $\tau_{i+1}$.
While for ideal hydrodynamics, this procedure works perfectly,   
  viscous hydrodynamics poses a problem
that shear stress tensor components contains 
time derivatives, $\partial_\tau \gamma_\perp$, $\partial_\tau u^x$,
$\partial_\tau u^x$ etc. Thus at the time step $\tau_i$ one needs the  
still unknown time derivatives.  
Numerically, time derivatives at step $\tau_i$ could be obtained
if velocities at time step $\tau_i$ and $\tau_{i+1}$ are known.
One possible way to circumvent the problem, is to use time derivatives of the previous step, i.e. use velocities at time step $\tau_{i-1}$ and $\tau_i$ to calculate the derivatives at time step $\tau_i$ \cite{Chaudhuri:2005ea}. It implicitly assume that the fluid velocity at the time step $i$ is the average of the fluid velocity at time step $i-1$ and $i+1$.

\section{Equation of state, viscosity coefficient and initial conditions} \label{sec4}

\subsection{Equation of state}

One of the most important inputs of a hydrodynamic model is  
the equation of state. Through this input, the macroscopic hydrodynamic models make contact 
with the microscopic world. 
In the present calculation we have used the equation of state,
EOS-Q, developed in ref.\cite{QGP3}.  It is a two-phase equation of
state. The hadronic phase of EOS-Q is modeled
as a non-interacting gas of hadronic resonance. As the temperature is increased,
larger and larger fraction of available energy goes into production of heavier and heavier resonances. This results into a
soft equation of state,
with small speed of sound, $c^2_s \approx 0.15$. With increasing temperature, the available volume is filled up with
resonances and the
hadronic states start to overlap, and microscopic degrees of freedom
are changed from   hadrons to deconfined quarks and gluons. The
QGP phase  is modeled as that of a 
non-interacting quark (u,d and s) and gluons, confined by a bag pressure
B. Corresponding equation of state, $p=\frac{1}{3}e -\frac{4}{3}B$ is 
stiff with a speed of sound $c_s^2 =\frac{1}{3}$. The two phases 
are matched by Maxwell construction at the critical temperature, $T_c=164 MeV$, adjusting the 
Bag pressure $B^{1/4}$=230 MeV. As discussed in \cite{QGP3}, ideal
hydrodynamics with the equation of state EOS-Q, explain a large volume of RHIC Au+Au data.

\subsection{Shear viscosity coefficient}

Shear viscosity coefficient ($\eta$) of dense nuclear medium (QGP or 
resonance hadron gas) is quite uncertain. While in principle, shear viscosity coefficient can be calculated in kinetic theory, it is a complex problem, more so in the strong coupling limit. The problem is yet unsolved. Recently,  using the ADS/CFT correspondence
\cite{Policastro:2001yc,Policastro:2002se} it was conjectured that the ratio of shear viscosity over the entropy ratio is bounded form the lower side. 
In a strongly coupled gauze theory, N=4 SUSY YM, the shear viscosity can be evaluated
as $\eta=\frac{\pi}{8}N^2_cT^3$ and the    entropy density 
is given by $s=\frac{\pi^2}{2}N^2_cT^3$. Thus in the strongly coupled field theory,

\begin{equation}
\left ( \frac{\eta}{s} \right )_{ADS/CFT} \geq \frac{1}{4\pi}\approx0.08,
\end{equation}

ADS/CFT correspondence suggests a minimal viscosity for the strongly couple gauze theories.  
In the present paper, we have limited our study to the evolution of minimally viscous ($\eta/s=1/4\pi\approx 0.08$) QGP fluid.  We assume that throughout the evolution, the ratio $\eta/s$=0.08 is maintained. 

\subsection{Relaxation time}

Relaxation time ($\tau_\pi$) is another important parameter in viscous hydrodynamics. As seen in Eqs.\ref{eqc4a}-\ref{eqc6}, in the time scale $\tau_\pi$ shear stress tensor $\pi^{\mu\nu}$ relaxes
to 1st order values, $\pi^{\mu\nu}=2\eta\sigma^{\mu\nu}$. Instantaneous values of the shear stress tensor and the associated thermodynamical quantities, characterizing the fluid, will depend on the relaxation time. Again in principle, kinetic theory 
can be used to compute the relaxation time, but as with the viscosity, for a strongly coupled system, the complex problem is yet unsolved.
In the Boltzmann gas limit, one can compute the  relaxation time,

\begin{equation}
\tau_{\pi}=\frac{6\eta}{4p}\approx \frac{6}{T}\frac{\eta}{s}.
\end{equation}

\noindent  where $p$ is the pressure. In the present paper we have used this estimate for the relaxation time,
However, we will also investigate the relaxation time dependence of the fluid evolution and subsequent particle production.

\subsection{Initial conditions} \label{ini}  

Numerical integration of Eqs.\ref{eqb1}-\ref{eqb3} and \ref{eqc4a}-\ref{eqc6} require initial conditions, 
 e.g. energy density $\varepsilon(x,y)$,  velocity components $v_x(x,y)$ and $v_y(x,y)$, the shear stress tensor components, $\pi^{xx}(x,y)$, $\pi^{yy}(x,y)$ and $\pi^{xy}(x,y)$ at the initial time $\tau_i$.  
A priori, the parameters are unknown,
only way to fix them is to confront the theory with experimental data.  
 
As discussed earlier, ideal hydrodynamics has been very successful in explaining a large volume of data in RHIC 200AGeV Au+Au collisions \cite{QGP3}.  In \cite{QGP3}, the initial energy density  profile was   parameterized geometrically.
At an impact parameter $\vec{b}$, transverse distribution of
wounded nucleons $N_{WN}(x,y,\vec{b})$    and of binary NN collisions 
$N_{BC}(x,y,\vec{b})$ are calculated in a Glauber model. 
A collision at impact parameter $\vec{b}$ is assumed to contain 25\% hard scattering
(proportional to number of binary collisions) and 75\% soft scattering
(proportional to number of wounded nucleons).  
Transverse energy density 
profile at impact parameter $\vec{b}$ is then obtained as,

\begin{equation}\label{eqIV2}
\varepsilon(x,y,\vec{b})=\varepsilon_0(0.75\times N_{WN}(x,y,\vec{b})+0.25 \times N_{BC}(x,y,\vec{b}))
\end{equation}

Assuming that at the initial time $\tau_i$, fluid velocity is zero ($v_x(x,y)=v_y(x,y)=0$), the   parameter $\varepsilon_0$ and the initial equilibration time
$\tau_i$ are fixed to reproduce the 
experimental transverse momentum distribution of pions in central Au+Au collisions. 
STAR and PHENIX data are fitted to obtain initial
equilibrium time $\tau_i$=0.6 fm and central  entropy density of $s=110 fm^{-3}$. 
This corresponds to peak energy density of the
fluid as 30 $GeV/fm^3$, or 
initial peak temperature of 350 MeV.  

In the present demonstrative calculation, we fix the initial time, energy density and fluid velocity, to the values obtained in ideal dynamics, i.e. at $\tau_i$=0.6 fm, $s_{ini}$=110$fm^{-3}$, $v_x=v_y$=0.  In dissipative hydrodynamics, additionally, one need to specify the initial transverse profile for the three independent shear stress tensor components, $\pi^{xx}$, $\pi^{yy}$ and $\pi^{xy}$. 
One choice could be to assume zero value initially, $\pi^{xx}=\pi^{yy}=\pi^{xy}=0$. However, as it will be shown later,
viscous effects are enhanced if initially shear stress tensor is non-zero.
Since we are considering is boost-invariant evolution, one choice for non-zero shear stress tensor could be the boost-invariant value.  We then assume that at  the equilibration time $\tau_i$, the dissipative fluxes have attained their longitudinal  boost-invariant values. The independent shear stress tensor components at the initial time is then obtained as, 
 
\begin{eqnarray}
\pi^{xx}(x,y)=&&\frac{2\eta(x,y)}{3\tau_i},\\ 
\pi^{yy}(x,y)=&&\frac{2\eta(x,y)}{3\tau_i},\\
\pi^{xy}(x,y)=&&0. 
\end{eqnarray}

However, we do emphasize that the above choice is largely arbitrary.   The correct values for initial $\pi^{\mu\nu}$ can only be obtained by confronting the experimental data. 
 
\section{particle spectra}\label{sec5}

Solutions of Eqs.\ref{eqb1}-\ref{eqb3} and \ref{eqc4a}-\ref{eqc6}
give the space-time evolution of the QGP fluid. The information about the time evolution of thermodynamic quantities e.g. energy density, fluid velocity and
shear stress tensor,  need to be converted into particle spectra to make contact with experiments. To do so, we use the well known Cooper-Frye prescription \cite{Cooper:1974mv}.
In Cooper-Frye prescription, the particle distribution is obtained by convoluting the one body distribution function over the freeze-out surface,

 \begin{equation} \label{eq6_1}
E\frac{dN}{d^3p}= \frac{dN}{dy d^2p_T} =\int_\Sigma d\Sigma_\mu p^\mu f(x,p)
\end{equation}

\noindent where $d\Sigma_\mu$ is the freeze-out hyper-surface and $f(x,p)$ is the
one-body distribution function. Now in ideal dynamics, the fluid is in local equilibrium and the one body distribution function is well approximated by the equilibrium distribution function,

\begin{equation}\label{eq6_2}
f(x,p)=f^{eq}(x,p)=\frac{g}{2\pi^3}\frac{1}{exp[\beta(u_\mu p^\mu -\mu)] \pm 1}.
\end{equation} 

\noindent with inverse temperature $\beta=1/T$ and chemical potential $\mu$. $g$ is the degeneracy factor.
In  viscous dynamics on the other hand, the fluid is not in
equilibrium and $f(x,p)$ can not be approximated by the equilibrium distribution function $f^{eq}(x,p)$.
In a highly non-equilibrium system,   distribution
function $f(x,p)$ is unknown. 
If the system is slightly off-equilibrium,  then
it is possible to calculate correction to equilibrium distribution 
function due to   (small) non-equilibrium effects. Slightly
off-equilibrium distribution function can be  approximated  as,

\begin{equation} \label{eq6_3}
f^{neq}(x,p)=f^{eq}(x,p) [1+\phi(x,p)],
\end{equation}

\noindent  $\phi(x,p) << 1$ is the deviation from equilibrium distribution 
function  $f^{eq}$. With shear viscosity as the only dissipative forces,
$\phi(x,p)$ can be locally approximated by a quadratic function 
of 4-momentum,

\begin{equation} \label{eqv4}
\phi(x,p)=\varepsilon_{\mu\nu} p^\mu p^\nu.
\end{equation}

Without any loss of generality $\varepsilon_{\mu\nu}$ can be written as
as,

\begin{equation} \label{eqv5}
\varepsilon^{\mu\nu}=\frac{1}{2(\varepsilon+p)T^2} \pi^{\mu\nu},
\end{equation}

\noindent completely specifying the non-equilibrium distribution function. As expected, correction factor increases with increasing viscosity. We also note that non-equilibrium correction depend quadratically on particle momentum. 
The effect of dissipation is more on large momentum particles.  
With the non-equilibrium distribution function thus specified, it is now possible to
calculate the particle spectra from the freeze-out surface. 
In $(\tau,x,y,\eta_s)$ coordinates, the freeze-out surface is parameterized as,

\begin{equation}
\Sigma^\mu=(\tau_f(x,y)\cosh \eta_s, x, y, \tau_f(x,y) \sinh \eta_s),
\end{equation}

\noindent and the normal vector on the hyper surface is,

\begin{equation}
d\Sigma_\mu=(\cosh \eta_s, -\frac{\partial \tau_f}{\partial x_f}, 
                        -\frac{\partial \tau_f}{\partial y_f}, -\sinh \eta_s)
\tau_f dx dy d\eta_s
\end{equation}

At the fluid position $(\tau,x,y,\eta_s)$ the particle 4-momenta are parameterized as,

\begin{equation}
p^\mu=(m_T cosh (\eta_s-Y), p^x, p^y, m_T sinh (\eta_s-Y))
\end{equation}

The volume element $p^\mu d\Sigma_\mu$ become,

\begin{equation}
p^\mu d\Sigma_\mu=(m_T cosh(\eta-Y)-\vec{p}_T. \vec{\nabla}_T \tau_f) \tau_f dx dy d\eta
\end{equation}

Equilibrium distribution function involve the term $\frac{p^\mu u_\mu}{T}$ which can be evaluated as,

\begin{equation}
\frac{p^\mu u_\mu}{T}=\frac{\gamma(m_T cosh(\eta-Y) -\vec{v}_T.\vec{p}_T -\mu/\gamma)}{T}
\end{equation}

The non-equilibrium distribution function require the sum
$p^\mu p^\nu \pi_{\mu\nu}$,

\begin{equation}
p_\mu p_\nu \pi^{\mu\nu}=a_1 cosh^2 (\eta-Y) +a_2 cosh(\eta-Y) + a_3
\end{equation}

with
\begin{eqnarray}
a_1=&&m_T^2(\pi^{\tau \tau} +\tau^2 \pi^{\eta \eta})\\
a_2=&&-2m_T(p_x \pi^{\tau x} + p_y \pi^{\tau y})\\
a_3=&&p_x^2 \pi^{xx} +p_y^2 \pi^{yy} +2p_x p_y \pi^{xy} - m_T^2 \tau^2 \pi^{\eta \eta}
\end{eqnarray}

Inserting all the relevant formulas in Eq.\ref{eq6_1} and integrating over 
spatial rapidity one obtains,
 
\begin{equation}
\frac{dN}{dyd^2p_T} =\frac{dN^{eq}}{dyd^2p_T}+\frac{dN^{neq}}{dyd^2p_T}
\end{equation}

with,

\begin{widetext}
\begin{eqnarray} \label{eq5_12}
\frac{dN^{eq}}{dyd^2p_T}=\frac{g}{(2\pi)^3} 
\int dx dy \tau_f [m_T K_1(n\beta) - p_T \vec{\nabla}_T \tau_f K_0(n\beta)]\\
\frac{dN^{neq}}{dyd^2p_T}=\frac{g}{(2\pi)^3} 
\int dx dy \tau_f [m_T\{ {\frac{a_1}{4} K_3(n\beta)+\frac{a_2}{2} K_2(n\beta)+(\frac{3a_1}{4}+a_3)K_1(n\beta)
+\frac{a_2}{2} K_0(n\beta)} \} \nonumber\\
- \vec{p}_T. \vec{\nabla}_T \tau_f \{\frac{a_1}{2} K_2(n\beta)+
 a_2 K_1(n\beta)+( \frac{a_1}{2}+a_3)K_0(n\beta)\}]
\end{eqnarray}
\end{widetext}

\noindent where  $K_0$, $K_1$, $K_2$ and $K_3$ are the modified Bessel functions.  

It must be mentioned that   the non-equilibrium distribution function $f^{neq}(x,p)$ is obtained as a correction to the
equilibrium distribution function $f^{eq}(x,p)$.   It is implied that the non-equilibrium correction is small $\phi(x,p) << 1$. Consequently, non-equilibrium contribution $\frac{dN^{neq}}{dyd^2p_T}$ to the invariant distribution must be small compared to the equilibrium contribution $\frac{dN^{eq}}{dyd^2p_T}$. The ratio,

\begin{equation}
R=\frac{dN^{neq}}{dN^{eq}}=\frac
{\frac{dN^{neq}}{dyd^2p_T}}{\frac{dN^{eq}}{dyd^2p_T}},
\end{equation}

\noindent must be much less than 1. The condition limits the applicability of viscous dynamics. From Eqs.\ref{eqv4} and \ref{eqv5}, we observe that the non-equilibrium correction depend quadratically on the transverse momentum and also  
 on the thermodynamic variables (energy density or temperature and shear stress tensor) at the freeze-out surface. The $p_T$ range over which viscous dynamics remains applicable thus depend on the freeze-out condition.

We will also show results for elliptic flow $v_2$. It is defined as,

\begin{equation}
V_2=\frac
{\int_0^{2\pi} \frac{dN}{dyd^2p_T} \cos(2\phi) d\phi}
{\int_0^{2\pi} \frac{dN}{dyd^2p_T}  d\phi}
\end{equation}

Expanding to the 1st order, elliptic flow as a function of transverse momentum can be obtained as,

\begin{eqnarray}\label{eqV19}
v_2(p_T) =&& v_2^{eq}(p_T) \nonumber \\&&+ \left(-v_2^{eq}\frac{\int d\phi \frac{d^2N^{neq}}{p_T dp_Td\phi}}{\int d\phi \frac{d^2N^{eq}}{p_Tdp_Td\phi}}
+\frac{\int d\phi cos(2\phi) \frac{d^2N^{neq}}{p_T dp_T d\phi}}
{\int d\phi \frac{d^2N^{eq}}{p_T dp_T d\phi}} \right) \nonumber \\
 = && v_2^{eq}(p_T) + v_2^{corr}(p_T),
\end{eqnarray} 
 
\noindent where  $v_2^{eq}$ is the  equilibrium elliptic flow  and $v_2^{corr}$ is the correction due to the non-equilibrium effects. 
Non-equilibrium correction grow quadratically with $p_T$ (see Eq.\ref{eq5}). $v_2^{eq}$ on the other hand grow less than linearly with $p_T$.   An important conclusion can be reached from Eq.\ref{eqV19}: asymptotically, viscous hydrodynamics do not predict saturation of elliptic flow. 

\section{Stability and accuracy test for AZHYDRO-KOLKATA}\label{sec6}

The energy-momentum conservation equations \ref{eqb1}-\ref{eqb3}, and the relaxation equations
\ref{eqc5}-\ref{eqc7}
are solved simultaneously using the code, AZHYDRO-KOLKATA,  developed at the Cyclotron Centre, Kolkata.   AZHYDRO-KOLKATA was built upon the publicly available code
"AZHYDRO" \cite{AZHYDRO} where energy-momentum conservation equations for the ideal fluid, undergoing arbitrary transverse expansion and boost-invariant longitudinal motion are solved using the SHASTA-FCT algorithm. AZHYDRO is extensively tested and results are published. We made extensive changes in the code 
to introduce dissipation due to shear viscosity. We have checked that AZHYDRO-KOLKATA reproduces the AZHYDRO results for 
dissipation free fluid. 
Equation of motion in viscous hydrodynamics dynamics can not be solved analytically. One can not    compare numerical solutions with analytical results. 
Any numerical code has to live with certain uncertainty. One try to limit the  uncertainty to an acceptable value. The general procedure for checking a numerical code is as follow:
(i) results should be stable against change in integration step lengths, 
(ii) any symmetry in the system should not be destroyed due to numerical inaccuracy and
(iii) numerical inaccuracy should not lead to unphysical maxima or minima. We test our code by checking whether these principles are violated or not.  

\begin{figure}[ht]
\includegraphics[bb=40 278 507 755 
 ,width=0.5\linewidth,clip]{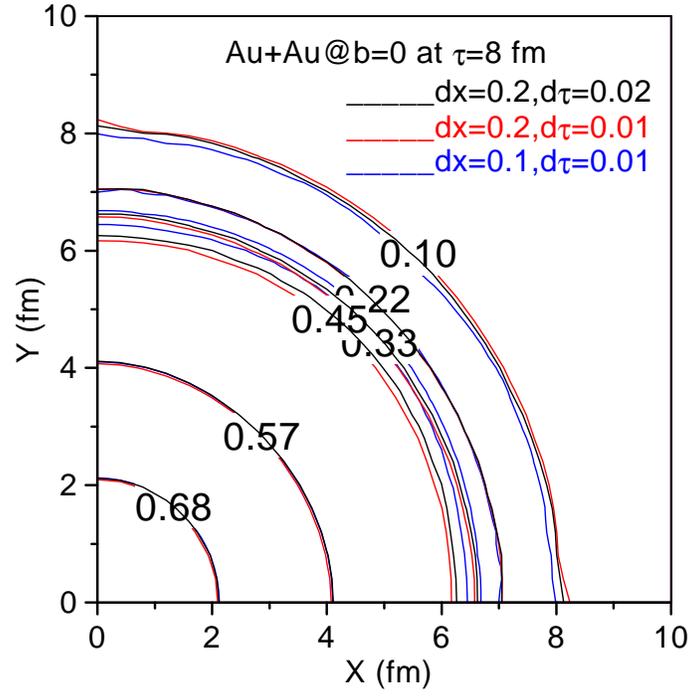}
\caption{\label{F1} (color online). Constant energy density contours in x-y plane after an evolution of 8fm. The black, red and blue lines are for integration step lengths: $dx=dy$=0.2 fm, $d\tau$=0.02 fm,
$dx=dy$=0.2 fm, $d\tau$=0.01 fm and $dx=dy$=0.1 fm, $d\tau$=0.01 fm. The viscosity coefficient is $\eta/s$=0.08.}
\end{figure}

In Fig.\ref{F1}, we have shown the contour plot of energy density
in x-y plane after an evolution of 8 fm in central Au+Au collision (b=0).  
 The black, red and blue lines are for integration step lengths, $dx=dy=0.2fm,d\tau=0.02 fm$,
$dx=dy=0.2fm,d\tau=0.01 fm$, and $dx=dy=0.1,d\tau=0.01 fm$ respectively. The
evolution of energy density is stable against change in integration step lengths. Even by halving the integration step lengths, the constant energy density 
surfaces occur nearly at the same spatial position. For example, spatial position of the 0.1 GeV surface is uncertain by less than a few percent.   In b=0 Au+Au collisions, the initial energy density distribution is symmetric with respect to x and y. Hydrodynamics evolution should not destroy the symmetry. Indeed, as seen in Fig.\ref{F1}, the x-y symmetry is maintained in the evolution. We also note that the energy density falls smoothly with the radius. There is no unphysical excess or deficit of energy density in the x-y plane. This indicate that the numerical solution of the hydrodynamic equations do not lead to any unphysical maxima or minima. 

\begin{figure}
\includegraphics[bb=38 278 507 748 
    ,width=0.5\linewidth,clip]{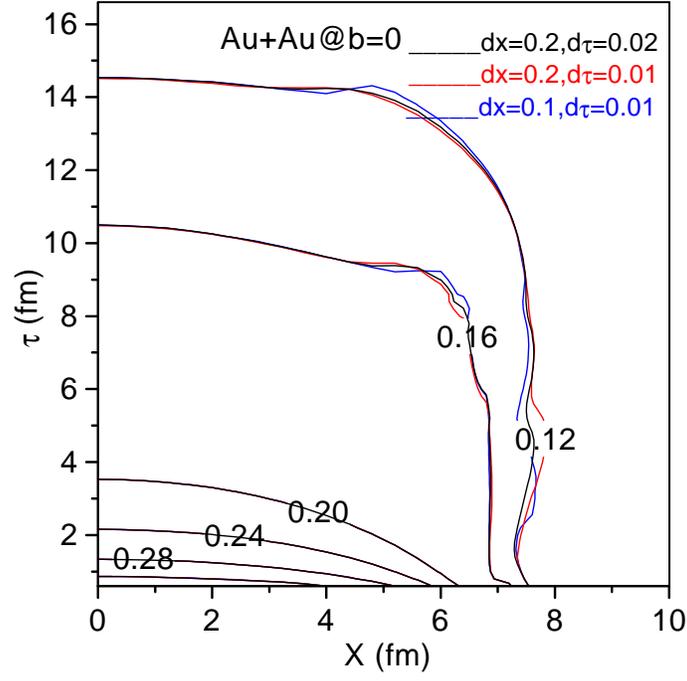}
\caption{\label{F2}
(color online).  constant temperature contours in $\tau-x$ plane. The black, red and blue lines are for integration step lengths, $dx=dy$=0.2 fm, $d\tau$=0.02 fm,
$dx=dy$=0.2 fm, $d\tau$=0.01 fm and $dx=dy$=0.1 fm, $d\tau$=0.01 fm respectively.
} 
\end{figure}

In Fig.\ref{F2}, the constant temperature contours in $x-\tau$ plane, at a fixed y=0 fm are shown. Constant temperature contour plots also show that the AZHYDRO-KOLKATA results are very stable against change in integration step lengths. 
Early in the evolution, the temperature evolution does not indicate any dependence on the integration step lengths. Only at later times,
that too at large distance from the center of the fluid, we find small 
change in temperature evolution. We are not showing but x and y components of the fluid velocity also show little dependence on the integration step lengths.

Evolution of shear stress tensor components also does not show any significant dependence on the integration step lengths. 
In Figs.\ref{F3} and \ref{F4},  we have shown the contour plot of the  $\pi^{xx}$  and $\pi^{yy}$  in $x-\tau$ plane, in an impact parameter b=6.5 fm Au+Au collision. The ordinate is at a fixed value, y=0. Here again, evolution of $\pi^{xx}$  and $\pi^{yy}$, with different integration step lengths, agrees with each other. It is expected also. Instantaneous values of energy density and velocity depend on the shear stress tensor. They are almost independent of integration step lengths. Naturally, shear stress tensor components do not show appreciable dependence on the integration step lengths. Figs.\ref{F1}-\ref{F4}, establish that the numerical solutions of the code, AZHYDRO-KOLKATA is stable against integration step lengths. It maintains the symmetry of the system and no unphysical maxima or minima   arise due to numerical inaccuracy.  

We have performed another check. 
Early in the evolution, even when the fluid is undergoing
arbitrary transverse and boost-invariant longitudinal expansion, at the center, the fluid is, least affected by the transverse motion. Early in the evolution, fluid at the center,  will follow the equation of motion for boost-invariant longitudinal expansion.  Causal viscous hydrodynamics for boost-invariant longitudinal expansion is well studied \cite{Muronga:2001zk}.  
For equation of state 
$p=1/3 \varepsilon=a T^4$ and viscosity $\eta=b T^3$, a and b are constant, temperature evolution of the fluid can be obtained by solving two ordinary differential equations \cite{Muronga:2001zk},
 
\begin{eqnarray} \label{eqV1}
\frac{dT}{d\tau}=&&-\frac{T}{3\tau} + \frac{T^{-3}\Phi}{12 a\tau}\\
\frac{d\Phi}{d\tau}=&&-\frac{2aT\Phi}{3b}-\frac{\Phi}{2}
\left( \frac{1}{\tau}-\frac{5}{T} \frac{dT}{d\tau}\right)+\frac{8aT^4}{9\tau} \label{eqV2}
\end{eqnarray}

\noindent where $\Phi=\tau^2 \pi^{\eta\eta}$.  

In Fig.\ref{F5}, we have compared the AZHYDRO-KOLKATA solutions at the center (x=y=0) with the solutions of one dimensional scaling expansion. The dotted lines are temperature evolution of fluid undergoing one dimensional scaling expansion. They are obtained by solving Eqs.\ref{eqV1} and \ref{eqV2}. We have shown results for two initial temperature $T_i$=0.450 GeV and 0.358 GeV, initial time is $\tau_i$=0.6 fm. The viscosity to entropy ratio is $\eta/s$=0.08. The solid lines in Fig.\ref{F5} are solutions from AZHYDRO-KOLKATA. They depict the temperature evolution of the fluid at the center.
For this particular plot, we have used Woods-Saxon type of distribution as the initial energy density distribution.   Woods-Saxon type of distribution is rather flat . Centre of the fluid will be less affected by the transverse expansion in initial Woods-Saxon type of distribution. 
Fig.\ref{F5} indicate that 
at the center, evolution of temperature of the fluid, undergoing
2+1 dimensional expansion, closely agree with the 
temperature evolution of the fluid in 0+1 dimension expansion.
The difference is less than 3\%.
Lastly, in Fig.\ref{F6} and \ref{F7}, we have shown the temperature evolution of the fluid in $x-\tau$ plane for $\eta/s$=0.08,0.04,0.02,0.01 and 0, respectively. In Fig.\ref{F6} and
\ref{F7} the contours are drawn for fixed y=0 fm and y=5 fm respectively. Ideal hydrodynamics results are recovered as $\eta/s$ gradually reduces to zero.
 The results shown in Figs.\ref{F1}-\ref{F7} give us confidence the code "AZHYDRO-KOLKATA" correctly evaluates the viscous hydrodynamics. 
 
In Fig.\ref{F1}-\ref{F7} 
we have used Israel-Stewart's relaxation equation (Eq.\ref{eq21}) to   compute the evolution of the fluid. As mentioned earlier, Eq.\ref{eq21} neglect the term $R=-[u^\mu\pi^{\nu\lambda}+u^\nu\pi^{\mu\lambda}]Du_\lambda$  under the approximation that both $\pi^{\mu\nu}$ and $Du_\mu$ are small and their product contribute only in 2nd order. It is important to check the accuracy of the approximation. We have simulated  b=0 and b=6.5 fm Au+Au collision with and without the term $R$ in the relaxation equation. Results are shown in Fig.\ref{F8} and \ref{F9}, where constant energy density contours in $x-\tau$ plane are drawn at a fixed value of y=0. The black and blue lines are for fluid evolution with and without the term R in the relaxation equation. Both for b=0 and b=6.5 fm Au+Au collisions, at early time energy density evolve nearly identically irrespective of the term R, indicating that the term R contribute minimally to fluid evolution. Only at late time, Israel-Stewart's relaxation equation results in marginally slower evolution.  
The result is understood. Initially, even though $\pi^{\mu\nu}$ is non-zero, the gradients of velocity is zero and the term R donot contribute. 
As the fluid evolve with time, velocity gradients grow but shear stress tensor $\pi^{\mu\nu}$ decreases. For minimally viscous fluid, velocity gradient do not grow to large value and the product  $\pi^{\mu\nu}Du_\mu$ contribute minimally.  

\section{Comparison of fluid evolution in ideal and minimally viscous fluid} \label{sec7}

In the following we compare the evolution of minimally  viscous ($\frac{\eta}{s}$=0.08)  fluid with ideal fluid evolution. 
For both the fluid, at the initial time $\tau_i$=0.6 fm, the peak entropy density is $s_{ini}$=110 $fm^{-3}$ with the Glauber model transverse density profile (see Eq.\ref{eqIV2}) and fluid velocity is zero, $v_x=v_y$=0.
The shear stress tensor is assumed to attain the boost-invariant value, $\pi^{xx}=\pi^{yy}=\frac{2\eta}{3\tau_i}$, $\pi^{xy}$=0.
In Fig. \ref{F10}, evolution of energy density in a b=0 Au+Au collision, in viscous and in ideal dynamics is compared. In the left panels (a), (b) and (c), the constant energy density contours in viscous dynamics after evolution of 2.6 fm, 4.6 fm and 8.6 fm are shown.
Their counterparts in ideal dynamics are shown in the right panels. Initially, energy density contours are identical both in ideal and in viscous simulations. As the fluid evolve, contour plots in ideal and viscous dynamics no longer remain identical. Evidently, energy density evolves slower in viscous dynamics. It is expected, viscosity opposes the expansion and cooling, the evolution is slowed down.

To understand better the evolution of the viscous vis. a vis. ideal fluid, in Fig.\ref{F11} and \ref{F12}, constant energy density contours in $x-\tau$ plane are drawn. The black and blue lines are for ideal and viscous fluid respectively. In Fig.\ref{F11}, contours are drawn at a fixed y=0 fm and in Fig.\ref{F12} at a fixed y=5 fm. Effect of viscosity is clearly seen, temperature evolution is slowed down. Viscous effect is comparatively large away from the center than near the center. It is understood. Velocity gradient generates viscosity. Velocity gradients are comparatively large away from the center. 
One also observes that transverse expansion seems to increase  with viscosity.

In viscous dynamics, fluid    cools 
slower that in ideal dynamics.  In Fig.\ref{F13}, temperature evolution of viscous and ideal fluids at a fixed y=0 fm, but at different values of x=0,1,2,4, and 6 fm are shown. The black and blue lines 
corresponds to ideal and viscous fluid evolution. Viscosity enhances the life time of the QGP phase, mixed phase. For a fixed value of $T_F$, life-time of the hadronic phase is also increased. 
It is obvious that in an inhomogeneous medium, effect of viscosity is not same everywhere. Viscous effects are comparatively less near the center of the fluid than at the periphery. 
 
Fluid velocity is assumed to be zero initially. With time fluid 
velocity grows. Evolution of fluid velocity in viscous and in ideal dynamics is compared in Figs.\ref{F14} and \ref{F15}.
Contours of constant $v_x$  and $v_y$ in x-y plane, after 8 fm of evolution is shown in Fig.\ref{F14} and Fig.\ref{F15}. The black and blue lines are for ideal and viscous fluid respectively. Contour plots indicate that in the interior of the fluid,   the fluid velocity grow
comparatively faster in viscous dynamics than in ideal hydrodynamics. 
In Fig.\ref{F16}, in four panels, we have shown the temporal growth of the x-component of the fluid velocity ($v_x$).  The black and blue lines are for ideal and viscous fluid. In the interior, fluid velocity starts to grow early. For example, at $x=y=1 fm$,
immediately after the evolution starts, the 
fluid velocity starts to grow. It grow for about 5 fm and then tends saturates.  Away from the center, growth is more rapid. Also the fluid velocity grows to larger value than at the interior. 
Fig.\ref{F16} indicate that in viscous dynamics, fluid velocity grow more rapidly than in ideal dynamics.    While in viscous dynamics, the fluid cools slower, the fluid velocity grow faster.

\subsection{Evolution of shear stress tensor}

As the fluid evolve, the shear stress tensor $\pi^{\mu\nu}$ also evolve. 
Temporal evolution of shear stress tensor components $\pi^{xx}$ and $\pi^{yy}$ is shown in Fig.\ref{F17}.  Contour plot of $\pi^{xx}$ and $\pi^{yy}$ in x-y plane, after evolution of 2.6 fm, 4.6 fm and 8.6 fm are drawn. We have initialized the shear stress tensor 
with boost-invariant values, $\pi^{xx}=\pi^{yy}=2\eta/\tau$.
$\pi^{xx}$ and $\pi^{yy}$ rapidly decreases with time. 
Initially, both $\pi^{xx}$ and $\pi^{yy}$ have similar distribution, but after a few fm of evolution,   the distribution of $\pi^{xx}$ and $\pi^{yy}$ starts to differ. One also observes that later in the evolution, $\pi^{xx}$ or $\pi^{yy}$ are stronger at the periphery than in the interior of the fluid. For example, after 4.6 fm of evolution, $\pi^{xx}$  in the interior are less by a factor of 5 or more than the value at the periphery. Velocity gradients generate shear stress tensor and they are strongest on the periphery. 
It is also interesting to note the similarity between the stress tensor components $\pi^{xx}$ and $\pi^{yy}$. As seen from the Eqs.\ref{eqc7} and \ref{eqc8}, $\pi^{xx}$ and $\pi^{yy}$ are related by a transformation $x\rightarrow y$ and $y\rightarrow x$. 
Fig.\ref{F17} correctly depict the relationship. 
We have not shown the evolution of the other independent component, $\pi^{xy}$. Initially, $\pi^{xy}$ is zero. With time $\pi^{xy}$ grow, but it do not grow to large value. $\pi^{xy}$ remains much less than the other two independent components $\pi^{xx}$ and $\pi^{yy}$.

\section{$p_T$ spectra and elliptic flow in viscous dynamics}\label{sec8}

In this section we will study the particle production in viscous dynamics. Space-time evolution of QGP fluid was solved using the computer code "AZHYDRO-KOLKATA". Assuming that the hadrons freezes out at a fixed temperature $T_F$, we construct the freeze-out surface and use the Cooper-Frey prescription to calculate the $p_T$ spectra and elliptic flow. We have ignored the resonance contribution to particle production.  

It was shown earlier that in AZHYDRO-KOLKATA, fluid evolution remains essentially unchanged if integration step lengths are halved. Nevertheless, fluid evolution does show small dependence on the integration step lengths. Let us first investigate the change in particle production when  equation of motions are integrated with step lengths, (i) $dx=dy$=0.2 fm, $d\tau$=0.02 fm and (ii) $dx=dy$=0.1 fm, $d\tau$=0.01 fm. 
With identical initial conditions, for Au+Au collisions at  b=6.5 fm,
we have solved minimally ($\eta/s$=0.08) viscous hydrodynamics, with the two sets of integration step lengths,   and computed pion yield from the freeze-out surface at $T_F$=150 MeV.  For the sake of comparison, we have also computed the yield in ideal dynamics, under similar conditions. In Fig.\ref{F18}(a) we have shown the $p_T$ distribution of $\pi^-$. The solid and the dashed lines correspond to ideal and viscous dynamics respectively. Black and blue lines show the yield obtained with  step lengths (i)$dx=dy$=0.2fm, $d\tau$=0.02fm and (ii)$dx=dy$=0.1fm,$d\tau$=0.01fm respectively.
  In ideal or viscous dynamics, transverse momentum distribution of $\pi^-$ do not show any significant dependence on integration step length. 
Whether the fluid evolution is calculated with step lengths 
 (i)$dx=dy=0.2fm,d\tau=0.02fm$  or (ii)$dx=dy=0.1fm,d\tau=0.01fm$,
over nine order of magnitude, pion yield   remain the same, 
The small variations the evolution with different integration step lengths, as observed in Fig.\ref{F1}-\ref{F4}, do not affect the pion transverse momentum distribution. 

In Fig.\ref{F18}(b), we have shown the elliptic flow in ideal and viscous dynamics. Elliptic flow is very sensitive observable. Unlike the $p_T$ distribution of $\pi^-$, both in ideal and viscous dynamics, elliptic flow of $\pi^-$ do show small dependence on the integration step lengths. By halving the step lengths, elliptic flow increases. However, the increase is small, less than 10\%.  The results shown in Fig.\ref{F18}b again confirm that numerical evaluation of viscous dynamics, as done in   AZHYDRO-KOLKATA is stable and reasonably accurate. 

Fig.\ref{F18} also illustrate the effect of viscosity on particle production. In viscous dynamics, $p_T$ spectra   is flattened. Compared to ideal dynamics, in viscous dynamics more particles are produced at large $p_T$. High $p_T$ particles are produced from the high temperature phase and  
in viscous dynamics, the fluid remains in the higher temperature phase for comparatively longer duration, enhancing the high $p_T$ yield. 
The elliptic flow on the other hand decreases in viscous dynamics. 
  Asymmetric pressure gradient generates elliptic flow. Apparently, in viscous dynamics, the shear stress tensor reduces the asymmetry in the pressure gradients. 
Indeed, appearance of additional pressure gradients  in the energy-momentum conservation Eqs.\ref{eqb2}-\ref{eqb3} does suggest that asymmetry in pressure gradients is reduced in viscous dynamics.  The increase of high $p_T$ yield and decrease of elliptic flow in viscous dynamics 
appears to remedy the drawbacks of the ideal hydrodynamics. It is known  that in ideal hydrodynamics, particle production at large $p_T$ is under-predicted while elliptic flow is over-predicted.

As mentioned earlier, viscous dynamics is applicable till the non-equilibrium contribution to  equilibrium production is less than unity. In Fig.\ref{F19}, the ratio of non-equilibrium to equilibrium contribution to pion yield, for freeze-out temperatures $T_F$=130,140,150 and 160 MeV, in  a b=6.5 fm Au+Au collision, are shown. The ratio depends sensitively on the freeze-out temperature. For freeze-out temperature $T_F$=160,150 and 140 MeV, non-equilibrium contribution exceed the 
equilibrium contribution beyond $p_T\approx$2.1, 2.5 and 3.5 GeV.
For $T_F$=130 MeV, the applicability range exceed 5 GeV.   The non-equilibrium correction depend on the thermodynamic variables on the freeze-out surface,  
$\varepsilon_{\mu\nu}=\frac{1}{2(\varepsilon+p)T^2} \pi^{\mu\nu}
\propto \pi^{\mu\nu}/T^6$ (assuming $\varepsilon=aT^4$) ,
(see Eq.\ref{eqv5}). The correction depend on two opposing effects, as the fluid freezes out at lower and lower temperature,
the correction increases due to $1/T^6$ dependence. However, 
as $T_F$ is lowered, the fluid evolve for longer time  and $\pi^{\mu\nu}$ decreases, the correction factor is decreased.
$\pi^{\mu\nu}$ decreases much faster than $1/T^6$ and the non-equilibrium correction is effectively reduced as the  freeze-out temperature is lowered. 

It is also interesting to note from Fig.\ref{F19} that at low $p_T$, the non-equilibrium correction makes a negative contribution to the equilibrium production. As similar effect is seen in \cite{Teaney:2003kp}, where, in a blast-wave model, effect of shear viscosity on the freeze-out surface was studied. Effect of viscosity on low $p_T$ and high $p_T$ particles are quite different. At low $p_T$, viscous effect
reduces the particle yield, while at high $p_T$, particle yield is increased.

\subsection{Comparison with experimental data}
\subsubsection{Elliptic flow}

Let us now confront the minimally viscous hydrodynamics with the experimental data in Au+Au collisions at RHIC.
Viscous hydrodynamics in 2+1 dimensions has a quite a few number of parameters,
(i) the initial  time ($\tau_i$), (ii) for a Glauber model type of transverse density profile, the peak initial energy density or equivalently the initial entropy density ($s_{ini}$) (iii) the initial fluid velocity ($v_x$ and $v_y)$ and (iv) the initial values for the three independent shear stress tensor components ($\pi^{xx}$, $\pi^{yy}$ and $\pi^{xy}$). Additionally, one require the freeze-out condition, the freeze-out temperature $T_F$. 
The shear viscosity and the relaxation time even though are calculable from kinetic theory, as stated earlier, the complex calculations are yet to be done and  presently, they has to be treated as parameters again. Very large number of parameters effectively reduces the efficacy of viscous dynamics. Better fit obtained to the data in viscous dynamics may be attributed to the increased parameter space. In the following, we limit our study to minimally viscous dynamics, $\eta/s$=0.08. We also fix the relaxation time to the Boltzmann gas estimate, $\tau_\pi=6\eta/sT$. Even then, the number of free parameters is large. We do not attempt to fit experimental data by varying all the parameters.  
As stated earlier, in ideal dynamics, if the QGP fluid is initialized at the initial time $\tau_i$=0.6 fm, to peak entropy density, $s_{ini}$=110 $fm^{-3}$,  and if the initial fluid velocity is zero, for freeze-out temperature $T_F$=100 MeV, the RHIC data on identified particle $p_T$ spectra and elliptic flow are reasonably well explained in a limited $p_T$ range, $p_T \leq $1.5 GeV \cite{QGP3}.  
To begin with we
assume that the initial conditions of the minimally viscous QGP fluid remains same as in ideal hydrodynamics, e.g. (i) $\tau_i$=0.6 fm, (ii) $s_{ini}$=110 $fm^{-3}$ (iii)$v_x=v_y$=0. 
 We also assume that at the initial time $\tau_i$,
 the  three independent shear stress tensor components have attained the boost-invariant values $\pi^{xx}=\pi^{yy}=2\eta/3\tau_i$, $\pi^{xy}$=0. 
We then try to fit the PHENIX data \cite{Adler:2004cj} on elliptic flow in 16-23\% Au+Au collisions, just by varying the freeze-out temperature ($T_F$). 
Being a ratio, elliptic flow   is very sensitive to the details of the model.   Hydrodynamics is better tested against the elliptic flow data than against the $p_T$ spectra.

Before we show the results of fit, it must be mentioned that,
since entropy is generated in viscous dynamics,  
assuming same initial condition for the viscous fluid as in ideal fluid will lead to increase in multiplicity. Indeed, we have checked that in central collisions, compared to ideal dynamics, $\pi^-$ multiplicity is increased by $\sim$ 28\% in viscous dynamics. In other word, if ideal dynamics with $\tau_i$=0.6 fm, $s_{ini}$=110 $fm^{-3}$ can reproduce the experimental multiplicity data, viscous dynamics, initialized similarly, will over-predict the data. However, since, ideal hydrodynamics, generally under-predict the multiplicity in central collisions (e.g. see Fig.5 in \cite{Kolb:2001qz}), viscous dynamics will possibly give better description of multiplicity data in central collisions. 

The solid circles in Fig.\ref{F20} are the PHENIX data on the transverse momentum dependence of the elliptic flow in 16-23\% Au+Au collisions.
16-23\% Au+Au collisions roughly corresponds to b=6.5 fm collision.
In Fig.\ref{F20}, in four panels,  for freeze-out temperatures, $T_F$=130,140,150 and 160 MeV the elliptic flow in  b=6.5 fm Au+Au collisions are shown. We have shown separately (i) the equilibrium contribution (the dash-dot line), (ii) the non-equilibrium contribution (the dash-dot-dot line) and (iii) the total contribution (the solid line).  The black arrows in the panels indicate the $p_T$ range above which the non-equilibrium contribution exceed the equilibrium contribution and viscous dynamics become inapplicable.  
  Several interesting points about elliptic flow in viscous hydrodynamics  can be noted. Up to $p_T$=2 GeV, the equilibrium elliptic flow ($v_2^{eq}$) is hardly affected by changing $T_F$. Only
beyond $p_T$=2 GeV, equilibrium elliptic flow increases with increasing $T_F$.  Even then the increase is marginal, e.g. at $p_T$=3 GeV, $v_2^{eq}$ increases by $\sim$ 10\% if $T_F$ is increased  from 130 to 160 MeV. One also note that the equilibrium flow agree 
with the PHENIX measurements up to $p_T\approx $ 3 GeV. 
If non-equilibrium correction to elliptic flow is ignored, it would have appeared that the PHENIX data on elliptic flow in 16-23\% centrality collisions is explained in the model for freeze-out temperature in the range 130-160 MeV.  
The non-equilibrium correction makes a negative contribution and reduces the elliptic flow.  
For $T_F$=140-160 MeV, the non-equilibrium contribution sufficiently reduces the elliptic flow and the total flow (equilibrium + correction) under-predict the PHENIX data. As the freeze-out temperature is lowered, magnitude of the non-equilibrium
correction decreases. For $T_F$=130 MeV, non-equilibrium correction to the equilibrium flow is very small and PHENIX data
are  well explained. As seen in Fig.\ref{F20},  data up to $p_T\approx$3.6 GeV are explained. 
 For comparison, in Fig.\ref{F20} we have shown the elliptic flow in ideal dynamics under identical conditions. In ideal dynamics, data are not explained beyond $p_T$=1.5 GeV.  

Even though, minimally viscous hydrodynamics describe the
elliptic flow in 16-23\% centrality Au+Au collisions, it appears that the
experimental saturation of elliptic flow at large $p_T$ is not explained.
Beyond $p_T$=3.6 GeV, elliptic flow continue to increase in viscous dynamics, though the rate of increase is slowed down. 
Saturation of elliptic flow is possibly beyond viscous hydrodynamics. Indeed, above $p_T$=3 GeV, "jet" physics become important. Jets can influence the elliptic flow. 
 Recently, in \cite{Chaudhuri:2007gq}, it was shown that elliptic flow is reduced in presence of jet quenching. 
Possibly viscous hydrodynamics, supplemented with jet physics will explain the saturation of elliptic flow at large $p_T$.
 
While minimally viscous hydrodynamics, for freeze-out temperature $T_F$=130 MeV,  give excellent description of elliptic flow in 16-23\% centrality collisions, in other centrality ranges of collisions, the agreement is not so good.
In Fig.\ref{F21},   PHENIX data \cite{Adare:2006ti}, on the elliptic flow in 0-10\%, 10-20\%, 20-30\% and 30-40\% centrality Au+Au collisions, are shown.  The solid lines in Fig.\ref{F21} are
viscous hydrodynamics predictions for
elliptic flow in    
b=3.2, 5.7, 7.4 and 8.7 fm Au+Au collisions. They  roughly corresponds to  0-10\%, 10-20\%, 20-30\% and 30-40\% Au+Au collisions. The freeze-out temperature is $T_F$=130 MeV.
Elliptic flow in 10-20\% and 20-30\% centrality collisions are
well explained in the model.  However the model under predict the
elliptic flow in  0-10\% centrality  collisions and over predict the flow in 30-40\% centrality collisions.  At $p_T$=2.3 GeV, viscous dynamics predicts $\sim$  35\% less elliptic flow in 0-10\% centrality collisions and $\sim$16\% more elliptic flow in 30-40\% centrality collisions. 
It appears that the minimally viscous dynamics explain the elliptic flow reasonably well only in mid-central collisions. In peripheral or in very central collisions, the predicted elliptic flow does not agree with the experiment. $p_T$ dependence of the minimum bias $v_2$ , on the other hand is excellently described in the model.
In Fig.\ref{F22}, the filled circles show the $p_T$ dependence of minimum bias $v_2$, as measured by the STAR collaboration \cite{Adams:2003zg}.
Data extends upto $p_T\approx$6 GeV. The solid line in Fig.\ref{F22} is the prediction from minimally viscous hydrodynamics. The experimental data are reproduced excellently.
Even though the model is incapable of explaining the elliptic flow in very central or very peripheral collisions, it does explain the minimum bias $v_2$.  As shown here, minimally viscous hydrodynamics under-predict $v_2$ in very central collisions and over-predict $v_2$ in peripheral collisions. The two opposite effects are neutralized on averaging and minimum bias $v_2$ is explained.   However,
we must remember that the present analysis is limited in the sense that the elliptic flow in 16-23\% centrality collisions was fitted by varying only the freeze-out temperature. The other parameters of the model e.g. initial time, initial energy density, initial shear stress tensor were kept fixed. Viscous evolution as well as particle production depends on those parameters.  It may be possible to explains the centrality dependence of elliptic flow, by tuning those parameters. Evidently, much more efforts will be needed to explain the centrality dependence of elliptic flow.

\subsubsection{Transverse momentum distribution}

Minimally viscous ($\eta/s$=0.08) QGP fluid, initialized to peak entropy density $s_{ini}$=110 $fm^{-3}$ at the initial time $\tau_i$=0.6 fm, if freezes-out at  temperature $T_F$=130 MeV, well reproduce the STAR data on the $p_T$ dependence of minimum bias $v_2$ as well as PHENIX data on $v_2$ in 10-20\%, 16-23\%, 20-30\% centrality Au+Au collisions. $v_2$
in very central or very peripheral collisions, is  not so well reproduced.  
Let us now confront the model predictions for $p_T$ spectra of identified particles with the experimental data. 
  In Fig.\ref{F23}, PHENIX data \cite{Adler:2003cb} on the transverse momentum distribution of $\pi^-$, in 0-5\%,5-10\%,10-20\%,20-30\%,30-40\% and 40-50\% centrality Au+Au collisions are shown.  Solid lines in Fig.\ref{F23} show the predictions from minimally  viscous hydrodynamics, in
Au+Au collisions at impact parameter b=2.3, 4.1, 5.7, 7.4, 8.7 and 9.9 fm. Roughly they corresponds to 0-5\%,5-10\%,10-20\%,20-30\%,30-40\% and 40-50\% centrality collisions. Model predictions are normalized by a factor $N=1.4$.
Minimally viscous hydrodynamics correctly reproduces the $p_T$ spectra of $\pi^-$ in all the centrality ranges of collisions. 
In contrast to ideal dynamics, where the transverse momentum spectra  of $\pi^-$ could be explained only up to $p_T\approx$1.5 GeV, in minimally viscous hydrodynamics, with the same initial conditions for the fluid, the spectra are explained right up to $p_T$=3 GeV.  Let us remind that, apart from the normalizing factor of N=1.4, there is no free parameter in the model. Considering that resonance production is neglected, the normalizing factor N=1.4 seems reasonable. 

Minimally viscous hydrodynamics also correctly reproduce the $p_T$ spectra of other identified particles e.g. $K^+$ and proton.
In Fig.\ref{F24} we have compared the minimally viscous hydrodynamic predictions for $p_T$ spectra of $K^+$ with the PHENIX data \cite{Adler:2003cb}.
PHENIX data on $K^+$ production cover a lower $p_T$ range, $p_T \approx 2 GeV$. In all the centrality ranges of collisions, the PHENIX data on $K^+$ are excellently reproduced in viscous dynamics. Here again, a comparable fit could not be obtained in ideal dynamics. Viscous hydrodynamics also well reproduces the
$p_T$ spectra of protons. In Fig.\ref{F25} predictions for proton spectra are compared
with the PHENIX experiment \cite{Adler:2003cb}.
PHENIX collaboration could measure proton spectra over much extended $p_T$ range, $p_T$=4.25 GeV. Minimally viscous hydrodynamics, explains the transverse momentum distribution of proton throughout the $p_T$ range. Here again, comparable description to the data could not be obtained in ideal hydrodynamics. 
 
Minimally viscous hydrodynamics also reproduces the 
centrality dependence of mean $p_T$ for $\pi^-$, $K^+$ and protons, in central and mid-central Au+Au collisions. In Fig.\ref{F26}, the PHENIX data \cite{Adler:2003cb} on the centrality dependence 
of mean  $p_T$ for $\pi^-$, $K^+$ and proton are shown. The
solid lines are viscous dynamics predictions. For $N_{part}\geq$=100,  $<p_T>$, for $\pi^-$, $K^+$ or protons are well reproduced. Only in peripheral collisions when $N_{part} < 100$, viscous dynamics produces more mean $p_T$ than in the experiment. 
Figures \ref{F20}-\ref{F26} clearly demonstrate the importance 
of viscosity. Even in minimally viscous dynamics, the viscous effect on particle production is considerable. Experimental data e.g. $p_T$ spectra of identified particles and elliptic flow, are much better explained in minimally viscous dynamics than in ideal dynamics. 

\section{Initial shear stress tensor and relaxation time dependence on fluid evolution and particle production} \label{sec9a}

In the foregoing analysis, we have assumed that initially, the shear stress tensor ($\pi^{\mu\nu}$) has attained the boost-invariant value,
$\pi^{xx}=\pi^{yy}=\frac{2\eta}{3\tau_i}, \pi^{xy}=0$. We have also 
used the Boltzmann gas estimate for the relaxation time ($\tau_\pi$).
Even though the choices seems to explain the PHENIX data on the centrality dependence of particle identified $p_T$ spectra, the centrality dependence of the elliptic flow is not satisfactory.
The fluid evolution and subsequent particle production depend on the initial $\pi^{\mu\nu}$ and also on $\tau_\pi$. They can be tuned to obtain better fit to the data. Even though,  we have not tuned them to fit the data, in this section, we study the dependence of fluid evolution and subsequent particle production on the initial value of $\pi^{\mu\nu}$ and the relaxation time $\tau_\pi$.

\subsection{Dependence of fluid evolution on initial shear stress tensor} 

 To investigate the dependence of the evolution, on the initial shear stress tensor $\pi^{\mu\nu}$, we have simulated  b=6.5 fm Au+Au collisions with (i)  boost-invariant values for the initial shear stress tensor, 
$\pi^{xx}=\pi^{yy}=2\eta/3\tau_i$, $\pi^{xy}$=0 and (ii)  zero values for the initial shear stress tensor, $\pi^{xx}=\pi^{yy}=\pi^{xy}$=0.
The other conditions of the fluid remaining the same, i.e. $\tau_i$=0.6 fm, $s_{ini}$=110 $fm^{-3}$, $v_x=v_y$=0, and $\tau_\pi=6\eta/sT$. In
Fig.\ref{F27} temporal evolution of the stress tensor component $\pi^{xx}$, for the two cases 
is compared. The blue and black lines are for initial $\pi^{xx}=0$ and $\pi^{xx}=2\eta/3\tau_i$ respectively.
The solid, dashed and medium dashed lines show the evolution at the fluid cell positions x=y=1,3 and 5 fm respectively. When initially $\pi^{xx}=0$, it grow rapidly to a maximum value   around $\tau$=1 fm.
Beyond $\tau$=1 fm, $\pi^{xx}$ decreases and after 5-6 fm of evolution, reduces to very small value. On the other hand, 
when initially $\pi^{xx}=2\eta/3\tau_i$, it decreases monotonically from the initial high value.  
A similar behavior is seen for the other independent component $\pi^{yy}$. From Fig.\ref{F27}, one understands that, other conditions remaining the same, viscous effects on the evolution is enhanced if, initially, shear stress tensor has attained the boost-invariant value rather than zero value.
It is also evident from Fig.\ref{F28}.
In Fig.\ref{F28}, the temperature evolution in the two cases, is compared. Constant temperature contours, in $x-\tau$ plane at a fixed y=0, are shown.   Fluid evolution is slower with initial non-zero initial shear stress tensor. 

 \subsection{Initial shear stress tensor dependence of $p_T$ spectra and elliptic flow}

As shown in Fig.\ref{F28}, effects of viscosity on fluid evolution is enhanced if $\pi^{\mu\nu}$ is initialized with the boost-invariant value rather than zero value. Since viscosity enhances high $p_T$ production and reduces the elliptic flow,
one expect more flattened $p_T$ spectra and less elliptic flow in evolution with initially boost-invariant $\pi^{\mu\nu}$ than in evolution with initially zero $\pi^{\mu\nu}$. In Fig.\ref{F29}a and b,
$p_T$ spectra and elliptic flow in a b=6.5 fm Au+Au collisions,
for (i) initial the boost-invariant $\pi^{\mu\nu}$ and (ii) initial zero $\pi^{\mu\nu}$, are shown.  Indeed, $p_T$ spectra is more flattened in evolution with initial boost-invariant $\pi^{\mu\nu}$ (see Fig.\ref{F29}a).  At $p_T$=3 GeV,
a factor of 5 increase in $\pi^-$ yield could be effected by changing the initial $\pi^{\mu\nu}$ from zero to boost-invariant value. In Fig.\ref{F29}b, $p_T$ dependence of elliptic flow is compared. Upto $p_T\sim$ 1.5 GeV, both the choice generate nearly same elliptic flow. Only beyond $p_T$=1.5 GeV, elliptic flow is less in evolution with initially boost-invariant $\pi^{\mu\nu}$ than in evolution with initially zero $\pi^{\mu\nu}$. 
Compared to evolution with initially zero $\pi^{\mu\nu}$,
at $p_T$=3 GeV, elliptic flow is reduced by $\sim$ 20\% in evolution with initially boost-invariant $\pi^{\mu\nu}$. 
In Fig.\ref{F29}b, we have also shown the PHENIX data on elliptic flow in 16-23\% centrality Au+Au collisions.  It appears that elliptic flow is better explained with initially boost-invariant $\pi^{\mu\nu}$. It is apparent that if initially shear stress tensor is zero, the data would require different freeze-out condition (or possibly different initial conditions).

\subsection{Dependence of fluid evolution on the relaxation time}

To investigate the effect of the relaxation time on fluid evolution, we have simulated Au+Au collision at b=6.5 fm,  with three different values of the relaxation time, (i)$\tau_\pi$=$\frac{3\eta}{sT}$, (ii) $\tau_\pi$=$\frac{6\eta}{sT}$ and (iii)$\tau_\pi$=$\frac{9\eta}{sT}$. The other conditions of the fluid remaining the same. Fluid evolution, for the three relaxation time, is compared in Fig.\ref{F30}.
 In Fig.\ref{F30} constant temperature contours in $x-\tau$ plane at a fixed y=0 fm, are drawn.  As the relaxation time increases, the fluid evolves slowly. Effect of relaxation time on the fluid evolution can be understood by examining the relaxation equation in one dimension,

\begin{equation}
\frac{d\pi}{d\tau}=-\frac{1}{\tau_{\pi}}(\pi-2\eta\sigma),
\end{equation}

\noindent the solution of which is $\pi(\tau)\approx 2\eta\sigma+exp(-\tau/\tau_{\pi})$.  
Larger the relaxation time, more is the instantaneous values of the shear stress tensor. Naturally effect of viscosity is comparatively large with large relaxation time and the fluid evolves slowly with increasing relaxation time.

\subsection{Relaxation time dependence of $p_T$ spectra and elliptic flow}
 
In Fig.\ref{F31}a and b, we have compared the $\pi^-$ $p_T$ spectra and elliptic flow in Au+Au collisions at b=6.5 fm, computed with three values of the relaxation time,
$\tau_\pi=3\eta/sT, 6\eta/sT$ and $9\eta/sT$. The other condition of the fluid remaining the same.
The $p_T$ spectra is flattened with increasing  relaxation time (see Fig.\ref{F31}a).   Particle production at high $p_T$ is increased if the stress tensor takes longer to relax. The increase could be substantial. For example, at $p_T$=3 GeV, changing relaxation time from $\frac{3\eta}{sT}$ to $\frac{9\eta}{sT}$, $\pi^-$ yield is increased nearly by a factor of 10.  In Fig.\ref{F31}b, we have shown the elliptic flow $v_2$.  Elliptic flow decreases as the relaxation time increases. For a factor of three increase in the relaxation time ($\tau_\pi=\frac{3\eta}{sT}$ to $\frac{9\eta}{sT}$), at $p_T\sim$ 3 GeV,  elliptic flow decreases by $\sim$ 15\%.  
In Fig.\ref{F31}b, for comparison sake, we  have shown the PHENIX data \cite{Adler:2004cj} on elliptic flow in 16-23\% centrality collision. It appears that Boltzmann gas value $\tau_\pi=6\eta/sT$, for the relaxation time best explain the PHENIX data on elliptic flow.

\section{Summary and conclusions} \label{sec9}

In Israel-Stewart's 2nd order theory of dissipative relativistic hydrodynamics, we have studied the evolution of QGP fluid in 2+1 dimensions. The "baryon free" QGP fluid (comprising u,d,s quarks and gluons) undergoes boost-invariant longitudinal motion and arbitrary transverse expansion. As the fluid expands its temperature decreases and when the temperature fall below a critical temperature $T_c$=164 MeV, the   fluid  undergoes a 1st order phase transition 
to hadronic fluid (comprising of all the hadronic resonances).

In 2nd order theory, in addition to the usual thermodynamic quantities e.g. energy density, pressure and hydrodynamic velocity, dissipative flows are treated as extended thermodynamic variables. Relaxation equations for the dissipative
flows are   solved, simultaneously with the energy-momentum conservation equations.
We have considered the most important dissipative effect, the shear viscosity. The bulk viscosity and heat conduction are neglected. 
ADS/CFT correspondence suggests a minimal viscosity for the QGP/hadronic fluid, $\eta/s$=0.08. 
We have limited our study to the evolution of minimally viscous QGP/hadronic  fluid.

In 2+1 dimensions, with boost-invariance,  only three shear stress tensor components are independent. We choose $\pi^{xx}$, $\pi^{yy}$ and $\pi^{xy}$ as the independent components. The dependent components are obtained from the constraints on the shear stress tensor, e.g. tracelessness and transversality to fluid velocity. The formulation ensures that throughout the evolution,
 the constraints on the shear stress tensor are satisfied.
The energy-momentum conservation equations and the relaxation equations for the independent shear stress tensor components, are solved
 using a computer
code "AZHYDRO-KOLKATA", developed at the Cyclotron Centre, Kolkata. The code was tested extensively to ascertain its stability and accuracy. 
Explicit simulation of ideal and viscous fluids, initialized under similar conditions  (e.g. same energy density, velocity profile at the equilibration time) confirms that the energy density or equivalently, the temperature  of a
viscous fluid, evolve slowly than its ideal counterpart.  
For a similar freeze-out condition, freeze-out surface is extended in
viscous dynamics. Fluid velocity on the other hand, develops faster in viscous dynamics. 

As the fluid evolve, shear pressure tensor $\pi^{\mu\nu}$ also evolve.
We have assumed that at the initial time $\tau_i$, $\pi^{\mu\nu}$ has attained the boost-invariant value.
As the fluid evolve, the independent components  $\pi^{xx}$ and $\pi^{yy}$, which are non-zero initially,
rapidly decrease and by 5-6 fm of evolution are reduced to very small values. The other independent shear stress tensor component $\pi^{xy}$ is zero initially. It grows with time, but never grow to large value.
Spatial distribution of $\pi^{xx}$ and $\pi^{yy}$ reveal an interesting feature of viscous dynamics. Even if, initially $\pi^{xx}$ and $\pi^{yy}$  have symmetric distribution, as the fluid evolve, pressure tensors quickly become asymmetric. However, in a central b=0, collision, we did not see any effect of the asymmetry in the energy density distribution. In a central
collision, the two opposite asymmetry cancels each other.  

In viscous hydrodynamics fluid evolution   depends considerably on the initial value of $\pi^{\mu\nu}$. If initially, instead of the boost-invariant value, $\pi^{\mu\nu}$ is zero,
effect of viscosity is much less on the evolution.   For boost-invariant value, initially,
$\pi^{xx}=\pi^{yy}=2\eta/\tau_i$, $\pi^{xy}$=0. As the fluid evolve, $\pi^{xx}$ and $\pi^{yy}$ rapidly decrease.
But in evolution with initial $\pi^{\mu\nu}=0$, within a time scale of $\sim$ 0.5 fm, 
$\pi^{xx}$ and $\pi^{yy}$ rapidly 
grow to reach a maximum and then decreases. The maximum reached is less than the corresponding boost-invariant value. Naturally, viscous effects are reduced. 
Viscous fluid evolution also depend on the
relaxation time. Less the relaxation time, more is the effect of viscosity on evolution.  For a given value of viscosity, viscous effect on the fluid evolution (and also on particle production) can be tuned by changing the relaxation time or the initial value of the shear stress tensor.  

We have also studied the effect of viscosity on particle production. Viscosity affects the particle production by (i) extending the freeze-out surface and (ii) by contributing a
non-equilibrium correction $\frac{dN^{neq}}{dyd^p_T}$ to the equilibrium particle $\frac{dN^{eq}}{dyd^2p_T}$ production. 
  The non-equilibrium correction must be small, $\frac{dN^{neq}}{dyd^p_T} << \frac{dN^{eq}}{dyd^2p_T}$. The non-equilibrium correction grow quadratically with $p_T$ and the condition $\frac{dN^{neq}}{dyd^p_T} << \frac{dN^{eq}}{dyd^2p_T}$ limits the $p_T$ range over which viscous hydrodynamics remains applicable. The non-equilibrium correction decreases as the fluid
freeze-out at lower and lower temperature. $p_T$ range of applicability also increases accordingly. Both the effects (i) and (ii),
enhances $p_T$ production and
under similar conditions, compared to ideal dynamics, in viscous 
dynamics, particle yield at large $p_T$ is increased. The elliptic flow on the other hand decreases.
Enhancement of $p_T$ production or reduction of elliptic flow in viscous dynamics can be understood in principle. 
In viscous dynamics, as the evolution is slowed down, the fluid spend more time in the higher temperature phase than in evolution
in ideal dynamics, and the high $p_T$ yield is increased.
In viscous dynamics, elliptic flow has two components, the equilibrium flow and the non-equilibrium correction to the equilibrium flow. The equilibrium flow itself is less than its ideal dynamics.
The non-equilibrium correction makes a negative contribution to further reduce the flow. If we remember that 
asymmetry in the pressure gradients generate the elliptic flow, apparently, in viscous dynamics, asymmetry in pressure gradients are reduced.  Indeed, appearance of additional pressure gradients in the energy-momentum conservation Eqs.\ref{eqb2}-\ref{eqb3} does suggest that asymmetry in the pressure gradients is reduced in viscous dynamics.

Fitting experimental data in  ideal hydrodynamics is a complex process, it is more complex in viscous hydrodynamics. As such ideal hydrodynamics has quite a few number of parameters, the initial time ($\tau_i$), the (transverse) energy density and velocity profile at the initial time and a freeze-out condition (the freeze-out temperature $T_F$).
In a remarkable effort, Kolb and Heinz \cite{QGP3} fitted the RHIC data on particle identified $p_T$ spectra and elliptic flow in Au+Au collisions. For the transverse density profile they used the Glauber model calculation (see Eq.\ref{eqIV2}). It was seen that 
ideal QGP fluid, thermalised at $\tau_i$=0.6 fm, with
peak entropy density $s_{ini}$=110 $fm^{-3}$ and zero fluid velocity, $v_x$=$v_y$=0, if freeze-out at a temperature $T_F$=100 MeV, explains the data in a limited $p_T$ range,   $p_T\approx$1.5 GeV. 
 
Parameter space is increased in viscous dynamics. As mentioned earlier, the three independent  shear stress tensor components ($\pi^{xx}$, $\pi^{yy}$ and $\pi^{xy}$) and the viscosity coefficient ($\eta$) and the relaxation time ($\tau_\pi$)  are to be treated as parameters. Since we consider only minimally viscous dynamics, the parameter $\eta$ is fixed, $\eta/s$=0.08.  We also fix the relaxation time to the  Boltzmann gas value, $\tau_\pi=6\eta/sT$. 
We also initialize the shear stress tensor $\pi^{\mu\nu}$ to boost-invariant value. 
To begin with we assume that in minimally viscous fluid, the initial condition of the fluid, i.e. equilibration time, initial entropy density, velocity,
will not be changed much from the value obtained in ideal dynamics and initialize the fluid as in ideal hydrodynamics, i.e. 
$\tau_i$=0.6 fm, $s_{ini}$=110 $fm^{-3}$, $v_x$=$v_y$=0. 
We only vary the freeze-out temperature ($T_F$) to fit PHENIX data \cite{Adler:2004cj} on the $p_T$ dependence of elliptic flow in 16-23\% centrality Au+Au collisions. The PHENIX data on the $p_T$ dependence of elliptic flow in 16-23\% Au+Au collisions can be well fitted if the hadronic fluid freeze-out at temperature $T_F$=130 MeV. Data up to $p_T\approx$3.6 GeV are then explained.
With the same freeze-out temperature i.e. $T_F$=130 MeV, PHENIX data \cite{Adare:2006ti} on the $p_T$ dependence of elliptic flow in mid central collisions e.g. 10-20\%, 20-30\% centrality collisions, are also well explained. The model also explains the STAR data \cite{Adams:2003zg} on the $p_T$ dependence of minimum 
bias $v_2$. However, the elliptic flow, in very central or peripheral collisions, is not so well explained. While in very central collisions the model under predict the elliptic flow,
in peripheral collisions, elliptic flow is over predicted. Apparently in very central collisions, PHENIX data on elliptic flow require less viscosity and in peripheral collisions data require more viscosity. 
Minimally viscous hydrodynamics with the parameters
fixed from elliptic flow in 16-23\% centrality Au+Au collisions, also explains the PHENIX data on $p_T$ spectra of identified particles.  $p_T$ spectra of $\pi^-$, $K^+$ and protons, in 0-10\%, 10-20\%, 20-30|5, 30-40\% and 40-50\% centrality collisions are explained more or less up to the experimental $p_T$ range. 
It is a great achievement of viscous dynamics. As mentioned earlier,
in ideal dynamics, data could be explained only up to $p_T\approx$1.5 GeV. The apparent contradiction i.e. minimally viscous hydrodynamics explains the centrality dependence of $p_T$ spectra of $\pi^-$, $K^+$ and proton, but fails the explain the elliptic flow in
very central or very peripheral collisions, 
indicate the sensitiveness of elliptic flow as compared to the $p_T$ spectra. 
Elliptic flow is a ratio and more sensitive to the details of the model than the $p_T$ spectra. 
Small difference between experiment and theoretical calculation, while unnoticed in the $p_T$ spectra gets amplified in  the elliptic flow. Centrality dependence of the elliptic flow tests the model more stringently than the centrality dependence of the $p_T$ spectra. Minimally viscous hydrodynamics also reproduces the centrality dependence of mean transverse momentum of $\pi^-$, $K^+$ and protons in central and mid-central collisions (up to  $N_{part}$=100).

To conclude, present analysis demonstrated the importance of shear viscosity in RHIC Au+Au collisions. Even with the ADS/CFT lower bound on viscosity, the $p_T$ spectra and elliptic flow are modified to a great extent and the experimental data are better explained than in ideal hydrodynamics. However, it will not be proper to claim that RHIC has produced fluid as ideal as possible.
A systematic analysis of RHIC data is required to check whether better, worse or equivalent description to the data are obtained with $\eta/s > 0.08$. Only then one can  claim about the QGP viscosity. 



\eject
{\bf Figure Captions:}
  
Fig.1: (color online). Constant energy density contours in x-y plane after an evolution of 8fm. The black, red and blue lines are for integration step lengths: $dx=dy$=0.2 fm, $d\tau$=0.02 fm, $dx=dy$=0.2 fm, $d\tau$=0.01 fm and $dx=dy$=0.1 fm, $d\tau$=0.01 fm. The viscosity coefficient is $\eta/s$=0.08. 

Fig.2: (color online).  constant temperature contours in $\tau-x$ plane. The black, red and blue lines are for integration step lengths, $dx=dy$=0.2 fm, $d\tau$=0.02 fm, $dx=dy$=0.2 fm, $d\tau$=0.01 fm and $dx=dy$=0.1 fm, $d\tau$=0.01 fm respectively.
 
Fig.3: (color online). Contour plot of $\pi^{xx}$ in $x-\tau$ plane, for a fixed value of y=0 are shown. The viscosity coefficient is $\eta/s$=0.08. The black and blue lines are for integration step lengths, dx=dy=0.2 fm, $d\tau$=0.02 fm and dx=dy=0.1 fm, $d\tau$=0.01 fm respectively.  
 
\begin{figure}[ht]
\includegraphics[bb=54 261 536 755
 ,width=0.5\linewidth,clip]{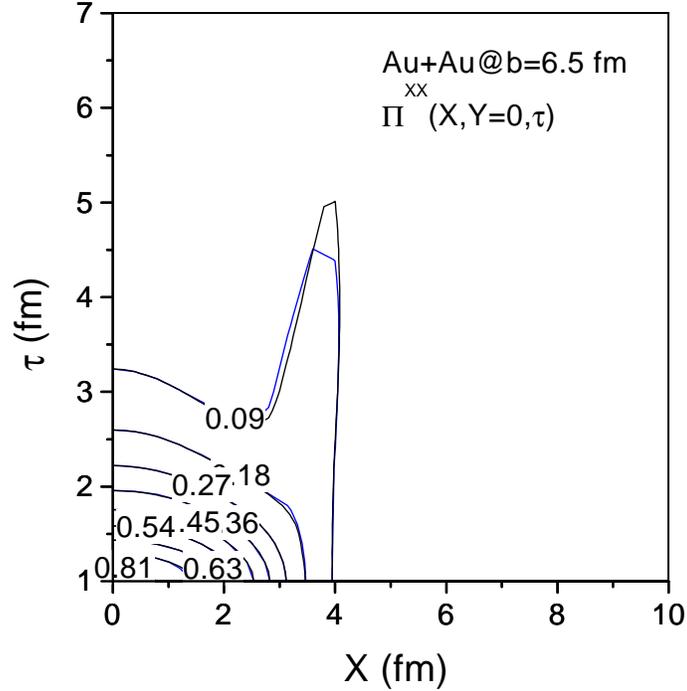}
\caption{\label{F3} (color online). Contour plot of $\pi^{xx}$ in $x-\tau$ plane, for a fixed value of y=0 are shown. The viscosity coefficient is $\eta/s$=0.08. The black and blue lines are for integration step lengths, dx=dy=0.2 fm, $d\tau$=0.02 fm and dx=dy=0.1 fm, $d\tau$=0.01 fm respectively.  }
\end{figure}

\begin{figure}
\includegraphics[bb=54 261 536 755
 ,width=0.5\linewidth,clip]{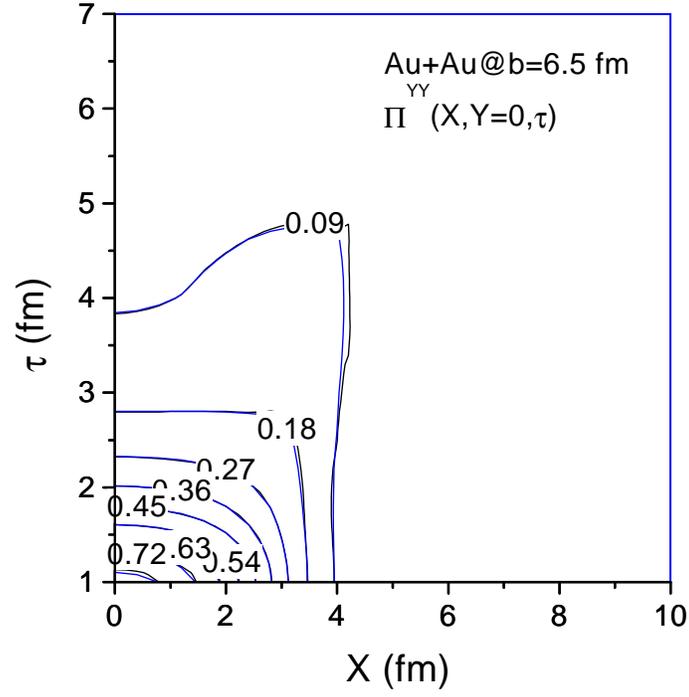}
\caption{\label{F4}(color online). Same as in Fig.\ref{F3} but for the shear stress tensor $\pi^{yy}$.  }
\end{figure}

\begin{figure} 
\includegraphics[bb=44 289 526 770
 ,width=0.5\linewidth,clip]{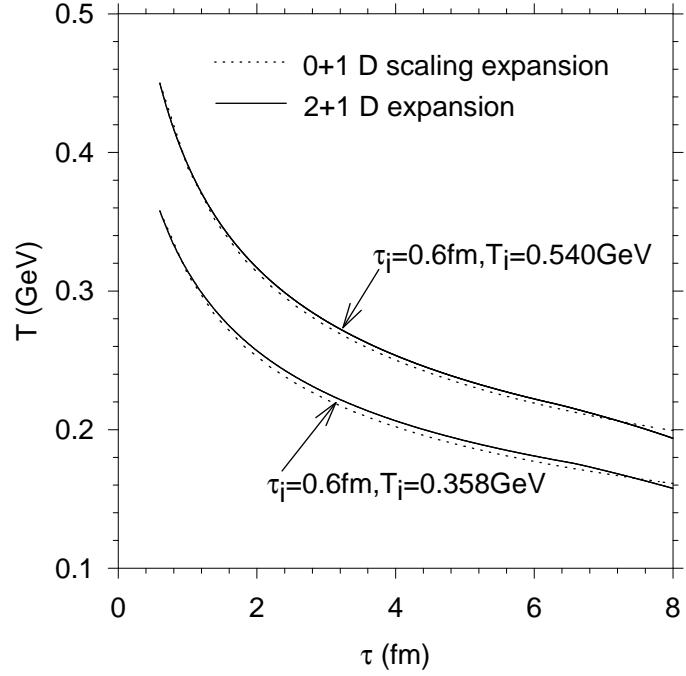}
\caption{\label{F5} The dotted lines depict the temperature evolution of viscous fluid in 0+1 dimension, for two initial temperature, $T_i$=.358 and .450 GeV. The solid lines are for the temperature evolution at the center of the fluid in 2+1 dimensions.}

\end{figure}

Fig.4: (color online). Same as in Fig.\ref{F3} but for the shear stress tensor $\pi^{yy}$.   

Fig.5: The dotted lines depict the temperature evolution of viscous fluid in 0+1 dimension, for two initial temperature, $T_i$=.358 and .450 GeV. The solid lines are for the temperature evolution at the center of the fluid in 2+1 dimensions. 

Fig.6: Contour plot of temperature in $x-\tau$ plane for fixed y=0. The five lines from top to bottom are for $\eta/s$=0.08,0.04,0.02,0.01 and 0 respectively. As the viscosity gradually reduces to zero, ideal hydrodynamics is recovered. 

\begin{figure}[ht]
\includegraphics[bb=50 246 526 735
 ,width=0.5\linewidth,clip]{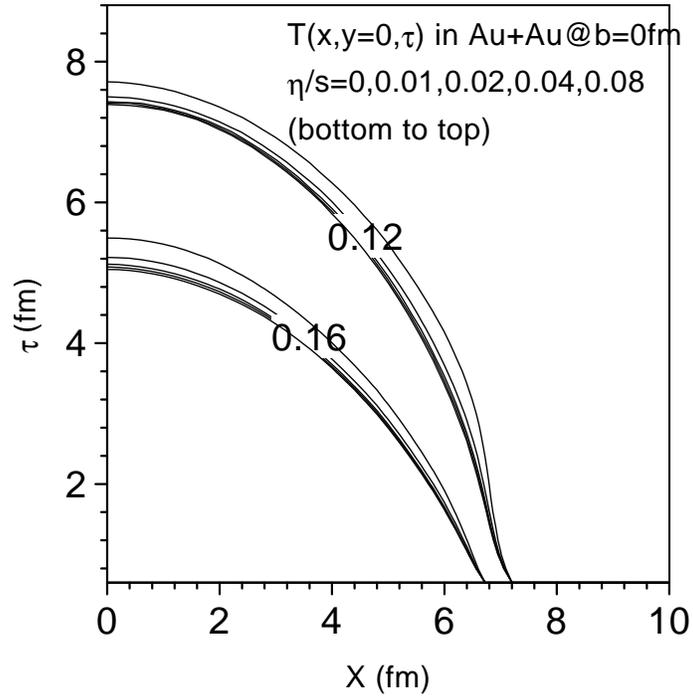}
\caption{\label{F6} Contour plot of temperature in $x-\tau$ plane for fixed y=0. The five lines from top to bottom are for $\eta/s$=0.08,0.04,0.02,0.01 and 0 respectively. As the viscosity gradually reduces to zero, ideal hydrodynamics is recovered.}
\end{figure}

\begin{figure}
\includegraphics[bb=50 246 526 735
    ,width=0.5\linewidth,clip]{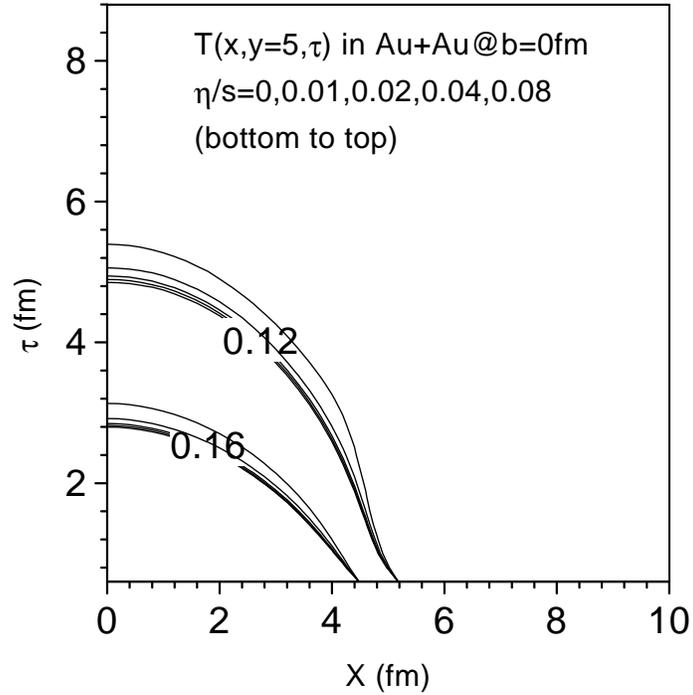}
\caption{\label{F7} Same as in \ref{F6}, but at y=5 fm.} 
\end{figure}

Fig.7: Same as in \ref{F6}, but at y=5 fm. 

Fig.8:(color online). Constant energy density contours in $x-\tau$ plane at a fixed value of y=0 in a b=0 fm Au+Au collision. The black lines are obtained with relaxation equation Eq.\ref{eqkt} containing the term $R$ and the blue lines are obtained when R is neglected (see text). Effect of term R is not large. 

\begin{figure}[ht]
\includegraphics[bb= 58 301 525 762
    ,width=0.5\linewidth,clip]{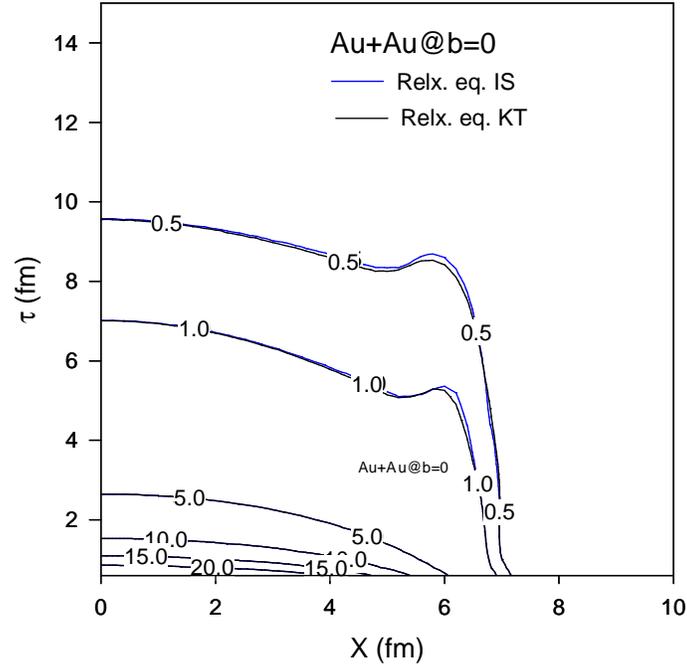}
\caption{\label{F8}
(color online). Constant energy density contours in $x-\tau$ plane at a fixed value of y=0 in a b=0 fm Au+Au collision. The black lines are obtained with relaxation equation Eq.\ref{eqkt} containing the term $R$ and the blue lines are obtained when R is neglected (see text). Effect of term R is not large.} 
\end{figure}

\begin{figure}[ht]
\includegraphics[bb= 60 268 535 743
    ,width=0.5\linewidth,clip]{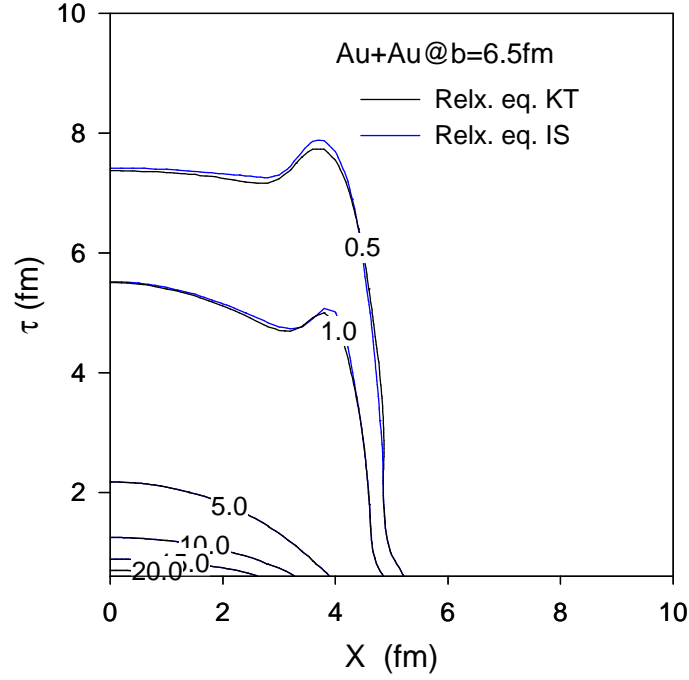}
\caption{\label{F9}(color online) same as in Fig.\ref{F8} but for  Au+Au collision at b=6.5 fm.} 
\end{figure}

\begin{figure}
\includegraphics[bb=32 203 551 808
 ,width=0.5\linewidth,clip]{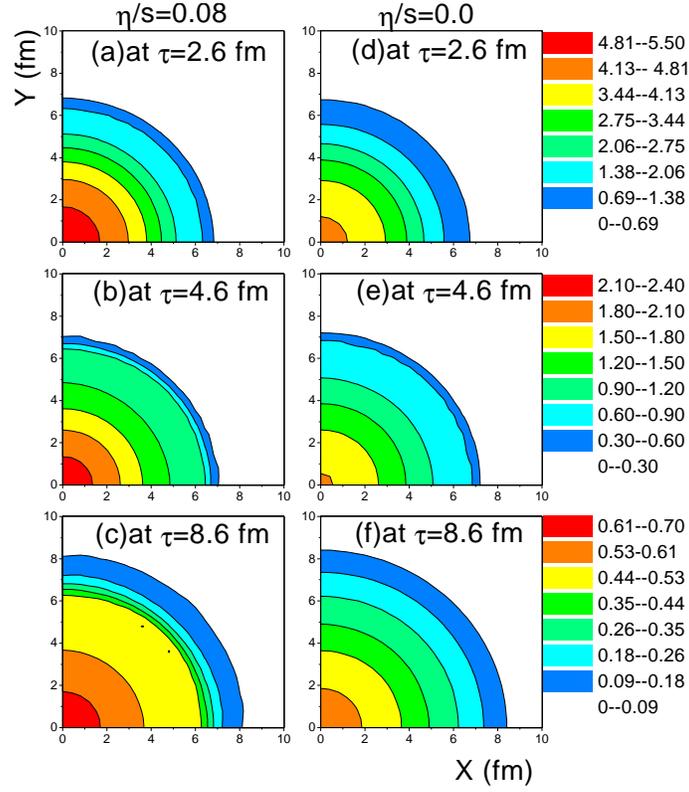}
\caption{\label{F10}(color online). Evolution of energy density in minimally viscous ($\eta/s$=0.08) and in ideal fluid evolution is compared. In panel a,b and c, constant energy density contours in x-y plane, at time $\tau$=2.6, 4.6 and 8.6 fm, in viscous evolution, are shown. Panels d,e and f show the energy density contours in ideal fluid evolution.}

\end{figure}

Fig.9:(color online) same as in Fig.\ref{F8} but for  Au+Au collision at b=6.5 fm. 

Fig.10:(color online). Evolution of energy density in minimally viscous ($\eta/s$=0.08) and in ideal fluid evolution is compared. In panel a,b and c, constant energy density contours in x-y plane, at time $\tau$=2.6, 4.6 and 8.6 fm, in viscous evolution, are shown. Panels d,e and f show the energy density contours in ideal fluid evolution. 

Fig.11:(color online). Constant energy density contours in $x-\tau$ plane at a fixed y=0 fm. The black and blue lines corresponds to ideal and viscous fluid evolution. 

\begin{figure}
\includegraphics[bb=38 254 560 755 
 ,width=0.5\linewidth,clip]{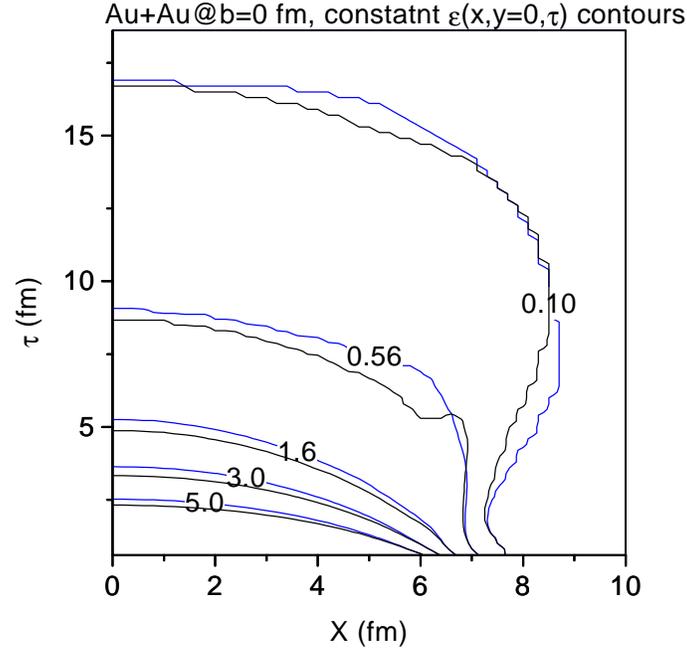}
\caption{\label{F11}
(color online). Constant energy density contours in $x-\tau$ plane at a fixed y=0 fm. The black and blue lines corresponds to ideal and viscous fluid evolution.}
\end{figure}

\begin{figure}
\includegraphics[bb=38 254 560 755
    ,width=0.5\linewidth,clip]{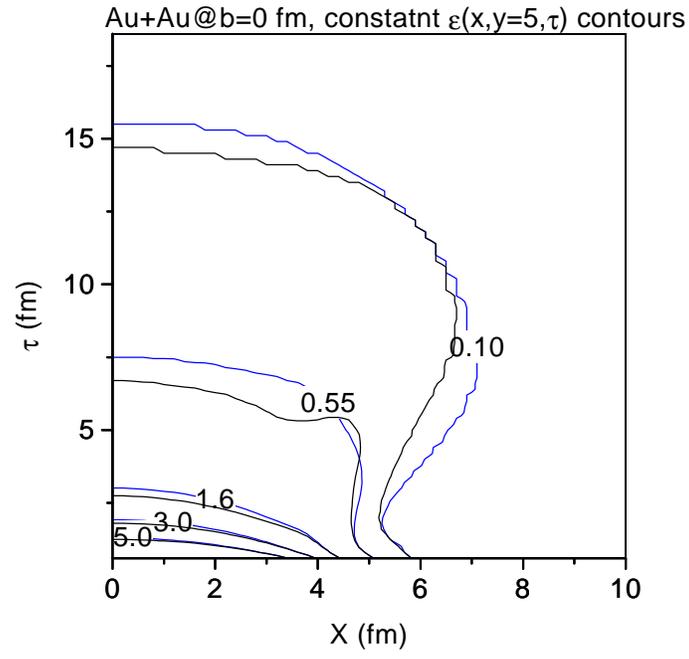}
\caption{\label{F12}
(color online). same as in Fig.\ref{F11}, but at y=5 fm.}
\end{figure}

\begin{figure}
\includegraphics[bb=44 290 527 770
 ,width=0.5\linewidth,clip]{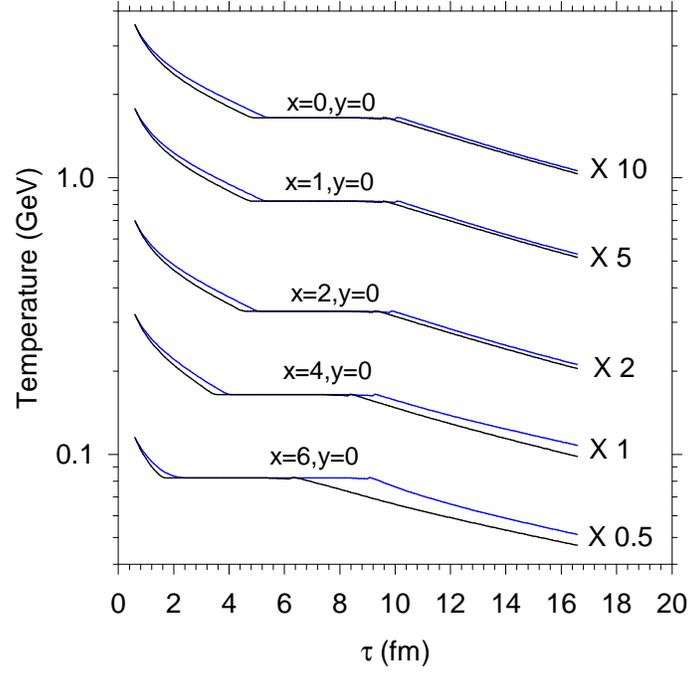}
\caption{\label{F13}(color online). Temporal evolution of fluid temperature at a fixed y=0 and different x-positions, x=0,1,2,4 and 6 fm are shown.   The black  and blue lines are for ideal and viscous fluid evolution. Multiplicative factor used for different curves are shown in the figure.}

\end{figure}

Fig.12: (color online). same as in Fig.\ref{F11}, but at y=5 fm. 

Fig.13: (color online). Temporal evolution of fluid temperature at a fixed y=0 and different x-positions, x=0,1,2,4 and 6 fm are shown.   The black  and blue lines are for ideal and viscous fluid evolution. Multiplicative factor used for different curves are shown in the figure. 

Fig.14:(color online).Constant $v_x$ contours in $x-y$ plane at $\tau$=8 fm. The black and blue lines correspond to ideal and minimally viscous fluid ($\eta/s$=0.08) evolution.    

\begin{figure}[ht]
\includegraphics[bb=26 267 508 768
 ,width=0.5\linewidth,clip]{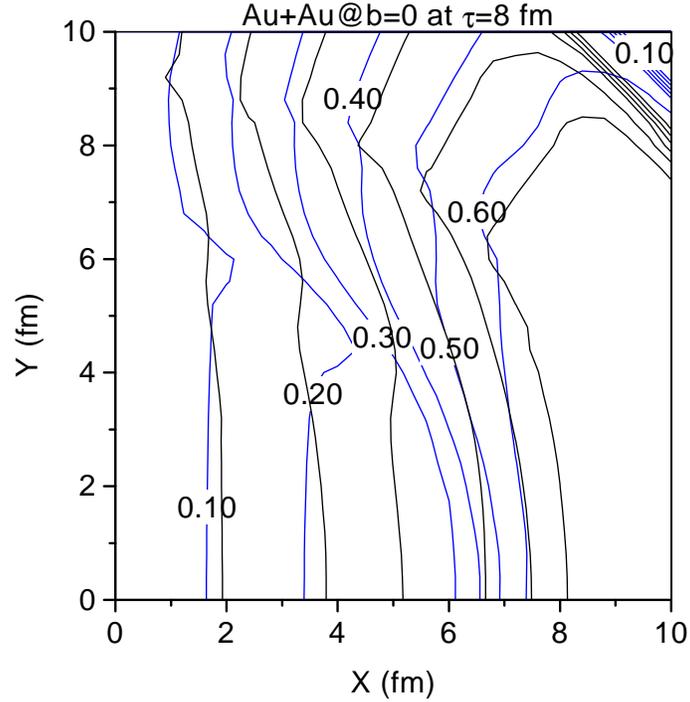}
\caption{\label{F14}
(color online).Constant $v_x$ contours in $x-y$ plane at $\tau$=8 fm. The black and blue lines correspond to ideal and minimally viscous fluid ($\eta/s$=0.08) evolution.    }
\end{figure}

\begin{figure}
\includegraphics[bb=25 268 507 770
    ,width=0.5\linewidth,clip]{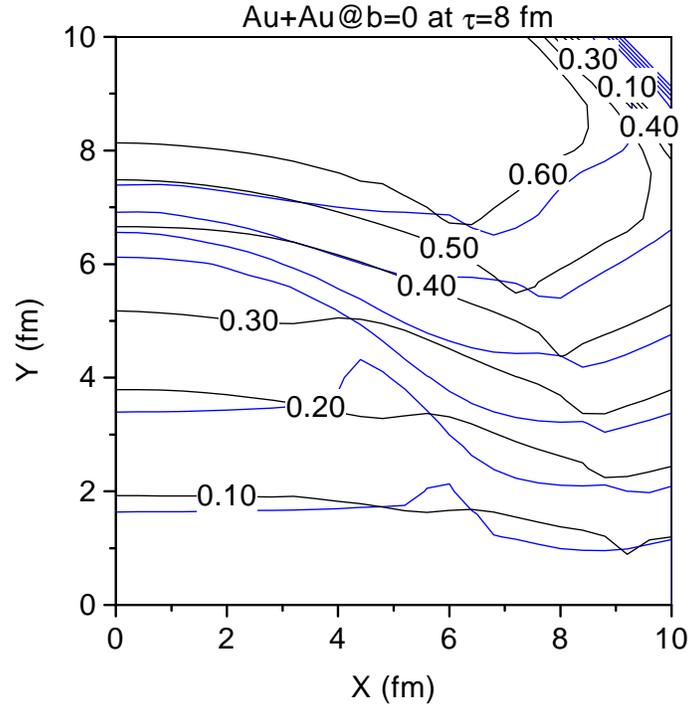}
\caption{\label{F15}
(color online). same as in Fig.\ref{F14} but for $v_y$. } 
\end{figure}

\begin{figure}
\includegraphics[bb=20 236 553 767
 ,width=0.5\linewidth,clip]{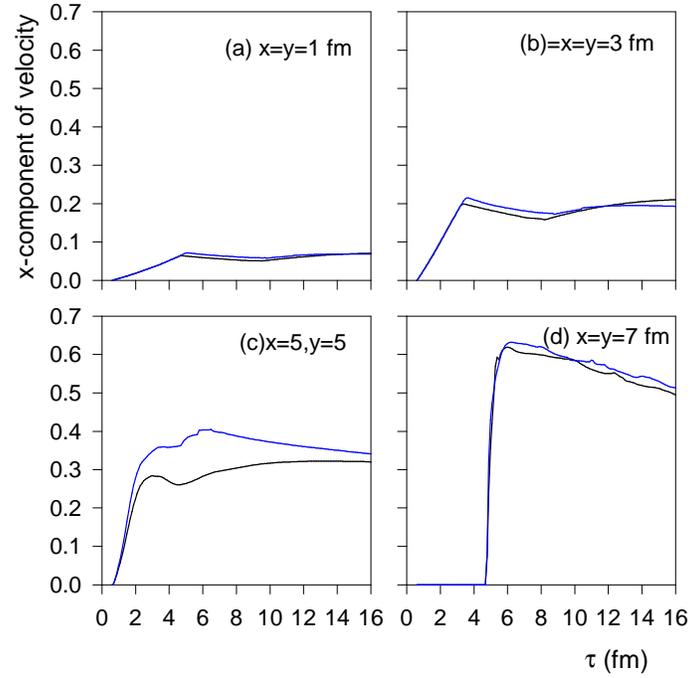}
\caption{\label{F16}(color online). Temporal evolution of the x-component of fluid velocity in ideal (the black lines) and in minimally viscous (the blue lines) dynamics is shown.  }
\end{figure}

\begin{figure}[ht]
\includegraphics[bb=13 14 581 827 
 ,width=0.5\linewidth,clip]{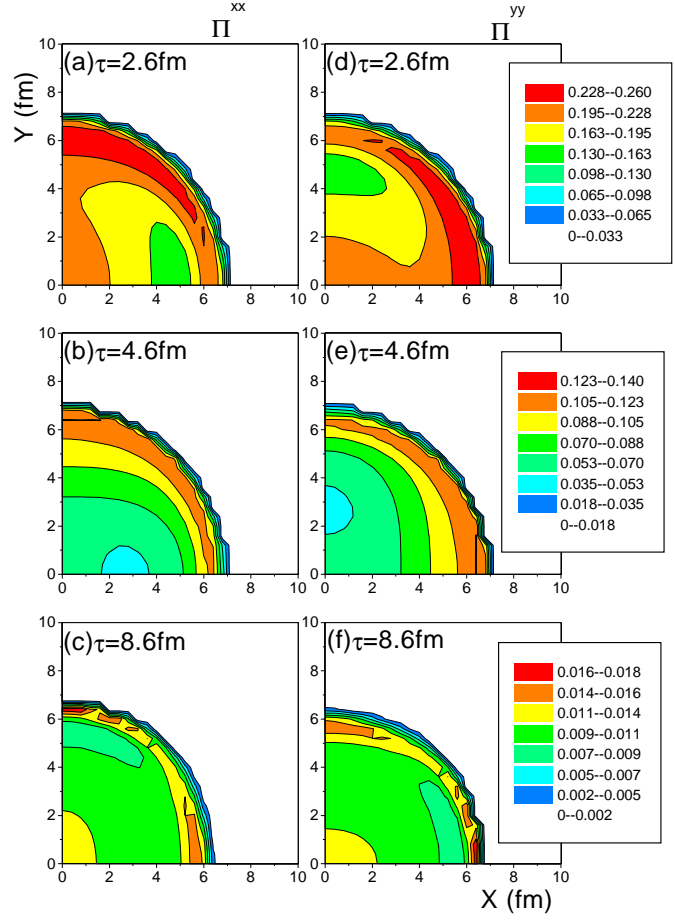}
\caption{\label{F17}(color online). In left panels (a), (b) and (c) contour plot
of the shear stress tensor component, $\pi^{xx}$  in $x-y$ plane at time $\tau$=2.6, 4.6 and 8.6 fm are shown. In right panels (d), (e) and (f) contours of constant $\pi^{yy}$ are shown.}
\end{figure}

\begin{figure}[ht]
\includegraphics[bb= 35 230 532 795
   ,width=0.5\linewidth,clip]{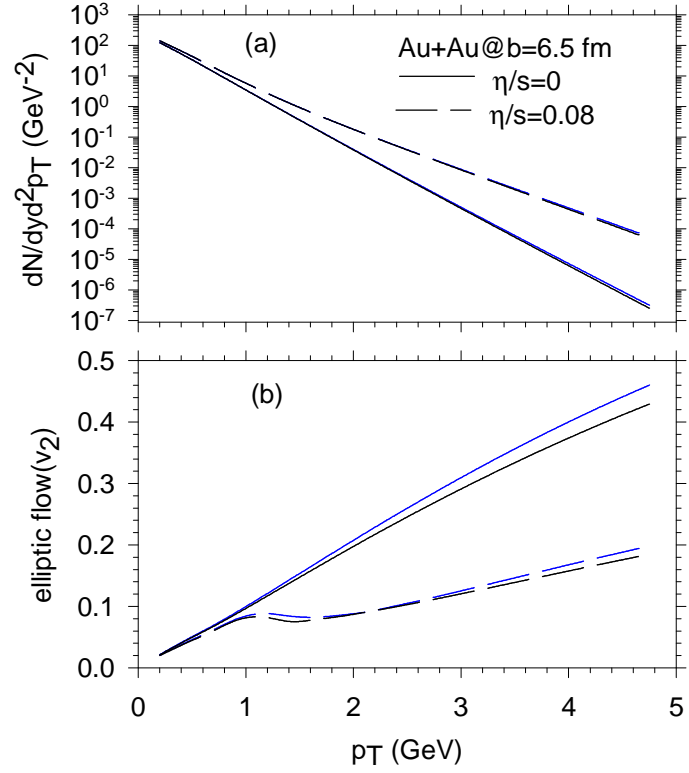}
\caption{\label{F18}(color online). (a)Transverse momentum spectra for $\pi^-$. The blue and black lines are obtained when viscous dynamics is solved with integration step lengths (i)dx=dy=0.02,$d\tau$=0.02 and (ii) dx=dy=0.01,$d\tau$=0.01 respectively. The solid lines and dashed lines corresponds to 
ideal and minimally viscous ($\eta/s$=0.08) fluid.( b) Same as in (a) but for the
elliptic flow.}
\end{figure}

\begin{figure}[ht]
\includegraphics[bb=37 290 524 769
   ,width=0.5\linewidth,clip]{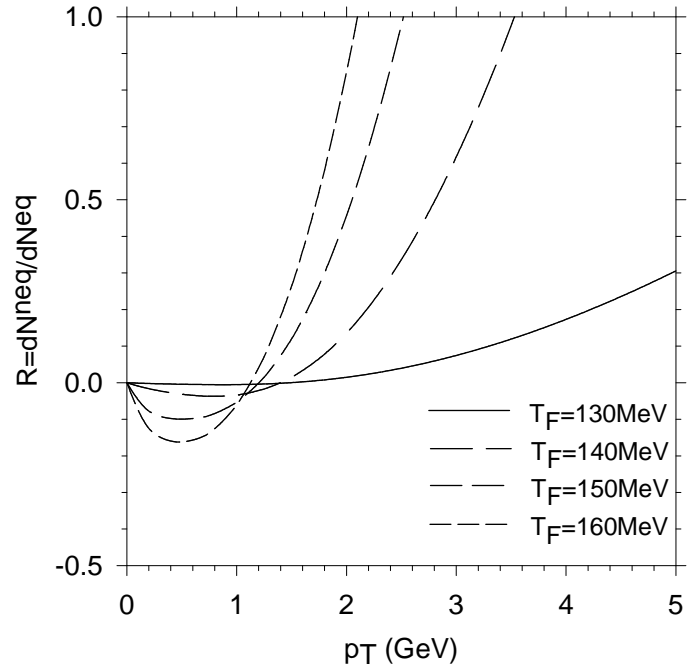}
\caption{\label{F19} The ratio of non-equilibrium contribution to equilibrium contribution to pion yield in b=6.5 fm Au+Au collisions. The solid, long dashed, medium dashed and short dashed lines are for freeze-out temperature $T_F$=130,140,150 and 160 MeV. For applicability of viscous dynamics, the ratios must be much less than unity.}
\end{figure}

\begin{figure} 
\includegraphics[bb=28 287 546 793
 ,width=0.5\linewidth,clip]{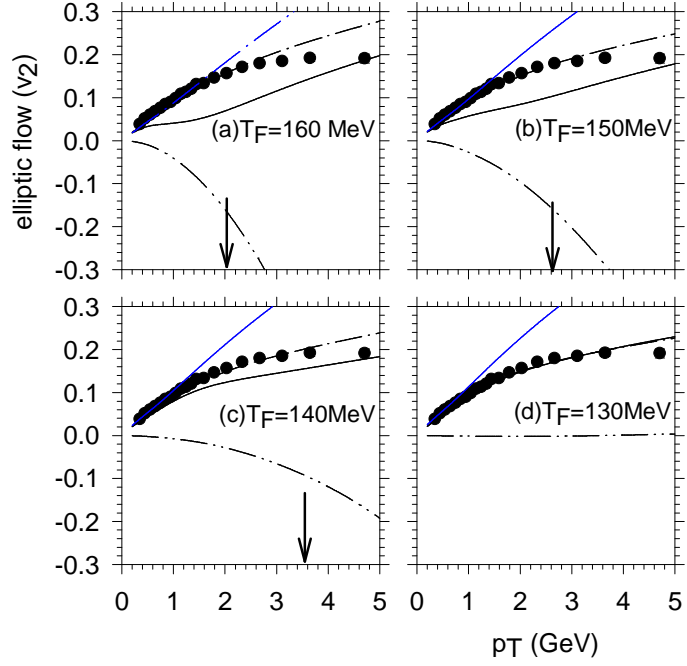}
\caption{\label{F20}
(color online). In four panels, $p_T$ dependence of elliptic flow in b=6.5 fm Au+Au collisions, for
freeze-out temperature $T_F$=160,150,140 and 130 MeV are shown. The dash-dot, dash-dot-dot and
the solid lines are equilibrium elliptic flow, the non-equilibrium correction to the equilibrium flow and the total flow (equilibrium+ non-equilibrium correction), in minimally viscous hydrodynamics. The blues lines are elliptic flow in ideal hydrodynamics under similar conditions. The filled circles are
the PHENIX data \cite{Adler:2004cj} on elliptic flow in 16-23\% centrality Au+Au collisions.} 
\end{figure}

\begin{figure}
 \includegraphics[bb=30 292 516 769
    ,width=0.5\linewidth,clip]{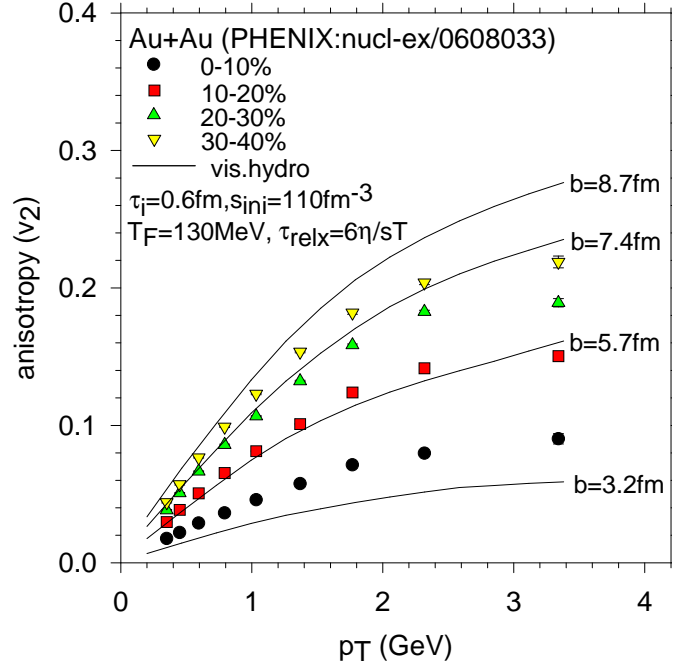}
\caption{\label{F21}
(color online).  PHENIX data \cite{Adare:2006ti} on the $p_T$ dependence of elliptic flow  
in 0-10\%, 10-20\%, 20-30\% and 30-40\% Au+Au collisions are shown. The solid lines are   predictions from minimally viscous hydrodynamics. }
\end{figure}

\begin{figure}  
\includegraphics[bb=46 291 524 769
    ,width=0.5\linewidth,clip]{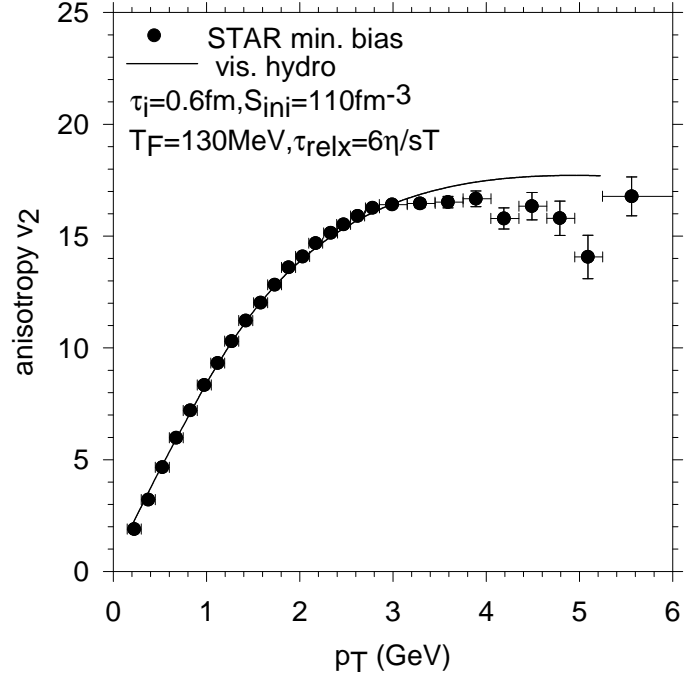}
\caption{ \label{F22}
   Filled circles are the STAR data \cite{Adams:2003zg} on the $p_T$ dependence of minimum bias elliptic flow in Au+Au collisions. The solid line is the  prediction from minimally viscous hydrodynamics. }
\end{figure}

Fig.15:(color online). same as in Fig.\ref{F14} but for $v_y$.  

Fig:16:(color online). Temporal evolution of the x-component of fluid velocity in ideal (the black lines) and in minimally viscous (the blue lines) dynamics is shown.   

Fig.17:(color online). In left panels (a), (b) and (c) contour plot
of the shear stress tensor component, $\pi^{xx}$  in $x-y$ plane at time $\tau$=2.6, 4.6 and 8.6 fm are shown. In right panels (d), (e) and (f) contours of constant $\pi^{yy}$ are shown. 

Fig.18:(color online). (a)Transverse momentum spectra for $\pi^-$. The blue and black lines are obtained when viscous dynamics is solved with integration step lengths (i)dx=dy=0.02,$d\tau$=0.02 and (ii) dx=dy=0.01,$d\tau$=0.01 respectively. The solid lines and dashed lines corresponds to ideal and minimally viscous ($\eta/s$=0.08) fluid.( b) Same as in (a) but for the
elliptic flow. 

Fig.19: The ratio of non-equilibrium contribution to equilibrium contribution to pion yield in b=6.5 fm Au+Au collisions. The solid, long dashed, medium dashed and short dashed lines are for freeze-out temperature $T_F$=130,140,150 and 160 MeV. For applicability of viscous dynamics, the ratios must be much less than unity. 

Fig.20:(color online). In four panels, $p_T$ dependence of elliptic flow in b=6.5 fm Au+Au collisions, for freeze-out temperature $T_F$=160,150,140 and 130 MeV are shown. The dash-dot, dash-dot-dot and the solid lines are equilibrium elliptic flow, the non-equilibrium correction to the equilibrium flow and the total flow (equilibrium+ non-equilibrium correction), in minimally viscous hydrodynamics. The blues lines are elliptic flow in ideal hydrodynamics under similar conditions. The filled circles are the PHENIX data \cite{Adler:2004cj} on elliptic flow in 16-23\% centrality Au+Au collisions. 

Fig.21:(color online).  PHENIX data \cite{Adare:2006ti} on the $p_T$ dependence of elliptic flow  in 0-10\%, 10-20\%, 20-30\% and 30-40\% Au+Au collisions are shown. The solid lines are   predictions from minimally viscous hydrodynamics.  

 Fig.22: Filled circles are the STAR data \cite{Adams:2003zg} on the $p_T$ dependence of minimum bias elliptic flow in Au+Au collisions. The solid line is the  prediction from minimally viscous hydrodynamics.  
   
Fig.23:(color online). PHENIX data \cite{Adler:2003cb} on $p_T$ spectra of $\pi^-$ in 0-5\%,5-10\%,10-20\%,20-30\%,30-40\% and 40-50\% centrality 
Au+Au collisions  are shown. The solid lines are predictions from minimally viscous hydrodynamics. 

\begin{figure}[ht]
\includegraphics[bb= 23 309 517 787
    ,width=0.5\linewidth,clip]{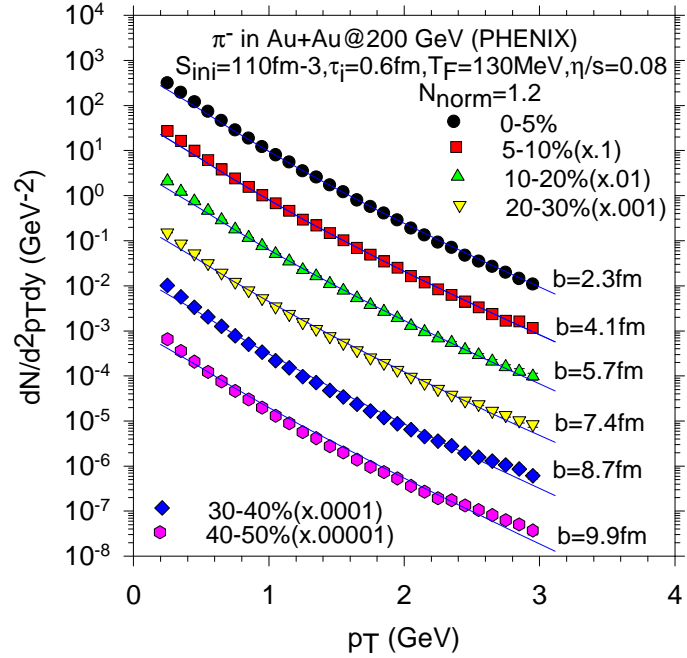}
\caption{\label{F23}
(color online). PHENIX data \cite{Adler:2003cb} on $p_T$ spectra of $\pi^-$ in 0-5\%,5-10\%,10-20\%,20-30\%,30-40\% and 40-50\% centrality 
Au+Au collisions  are shown. The solid lines are predictions from minimally viscous hydrodynamics.}
\end{figure}

\begin{figure} 
\includegraphics[bb=68 291 524 770
  ,width=0.5\linewidth,clip]{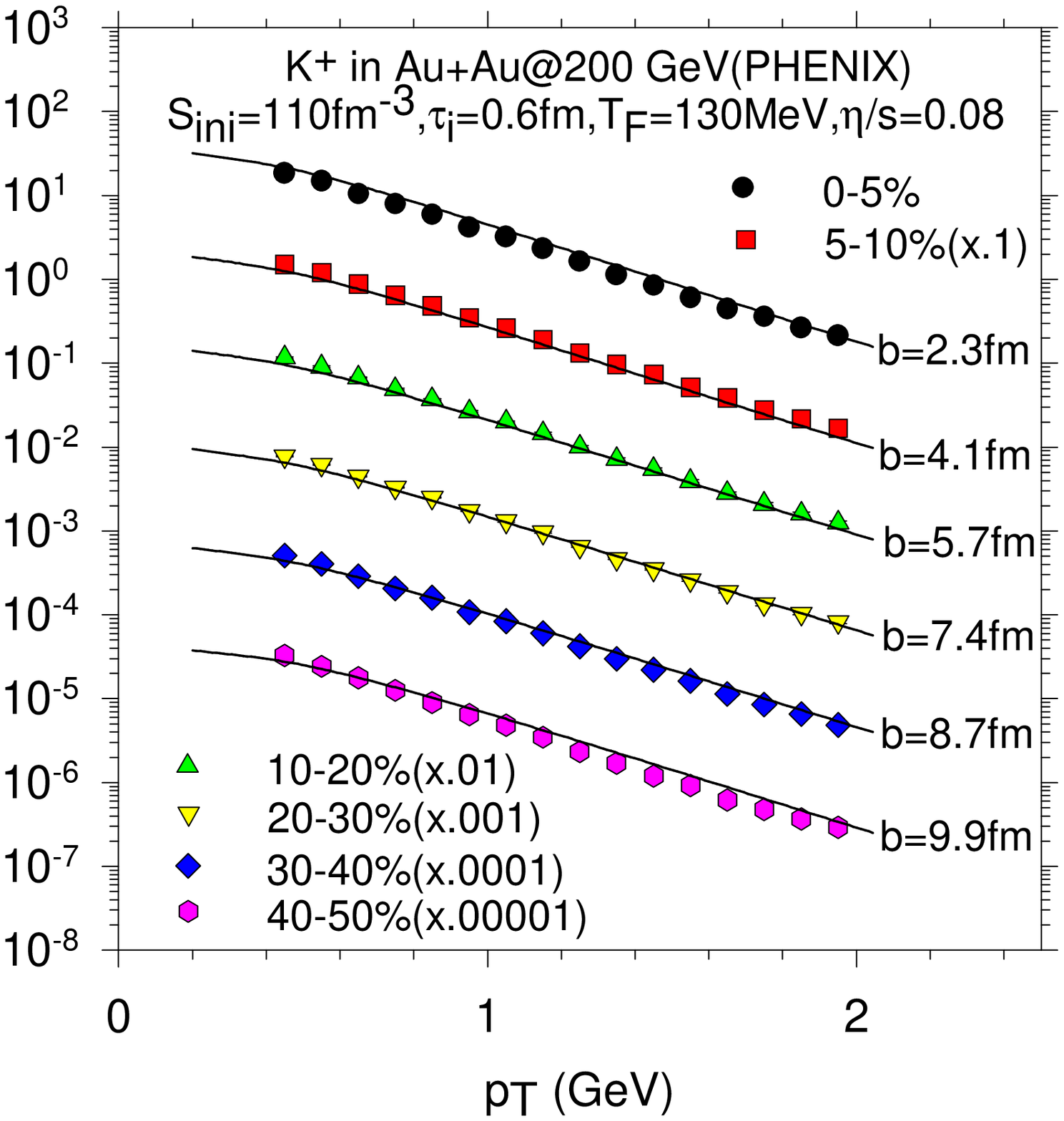}
\caption{\label{F24}
(color online). Same as in Fig.\ref{F23} but for $K^+$.}
\end{figure}

\begin{figure} 
\includegraphics[bb=22 291 524 768
    ,width=0.5\linewidth,clip]{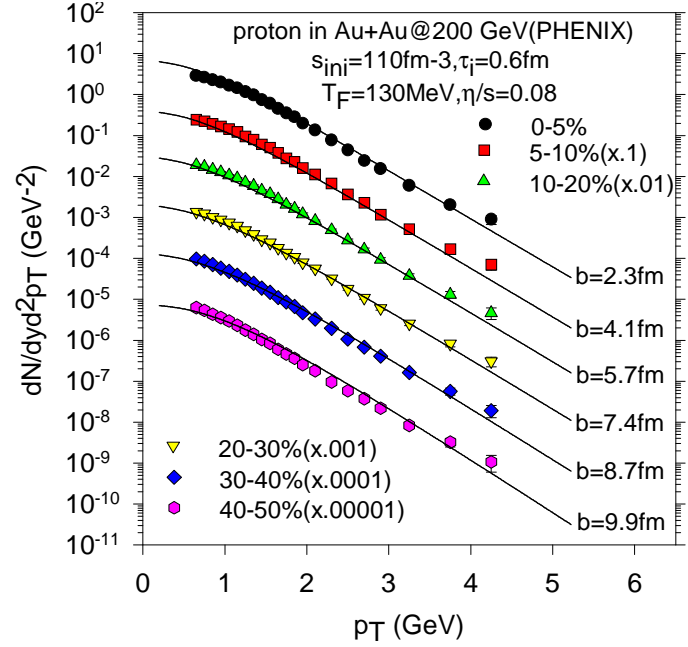}
\caption{\label{F25}
(color online). Same as in Fig.\ref{F23} but for proton.}
\end{figure}

\begin{figure} 
 \includegraphics[bb=40 287 533 769
 ,width=0.5\linewidth,clip]{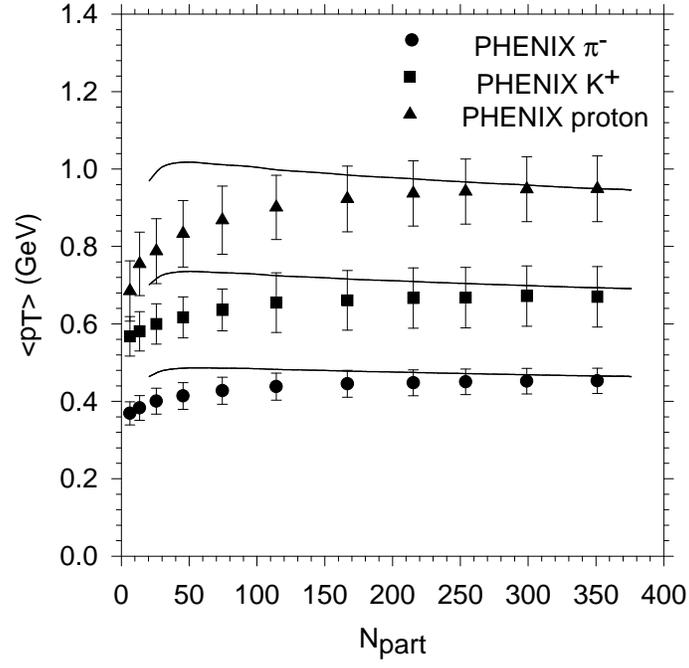}
\caption{\label{F26} PHENIX data on the centrality dependence of average $p_T$ for $\pi^-$, $K^+$ and protons are shown. The solid lines are predictions from minimally viscous dynamics.} 
\end{figure}

\begin{figure} 
\includegraphics[bb=38 289 526 770 
 ,width=0.5\linewidth,clip]{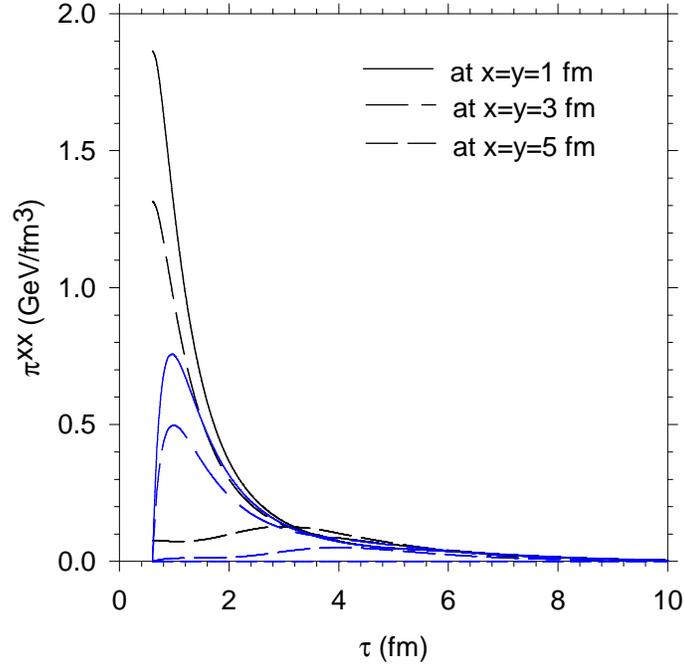}
\caption{\label{F27}  
(color online). Evolution of shear stress tensor $\pi^{xx}(x,y=0)$ for x=0,1,3 and 6 fm. The blue and black lines corresponds to fluid evolution with zero and non-zero (boost-invariant value) initial shear stress tensors.} 
 \end{figure}

\begin{figure} 
\includegraphics[bb=38 255 524 735
 ,width=0.5\linewidth,clip]{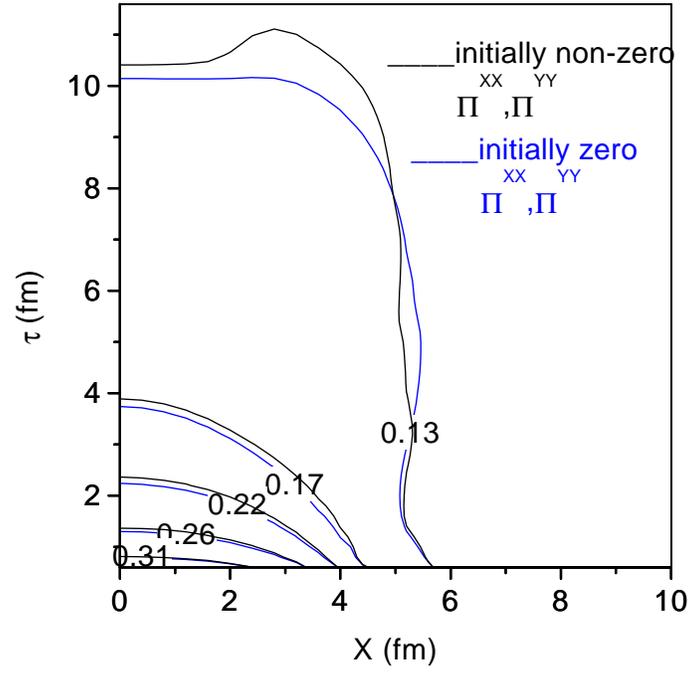}
\caption{\label{F28} (color online). Dependence of fluid evolution on initial shear stress tensor. Constant temperature contours  
in $x-\tau$ plane at a fixed y=0 fm, in a Au+Au collision at b=6.5 fm are shown. The black lines correspond to initially zero shear stress tensor. The blue lines correspond to initially non-zero,  boost-invariant shear stress tensors.   }
\end{figure}

\begin{figure} 
\includegraphics[bb=35 188 532 808 
 ,width=0.5\linewidth,clip]{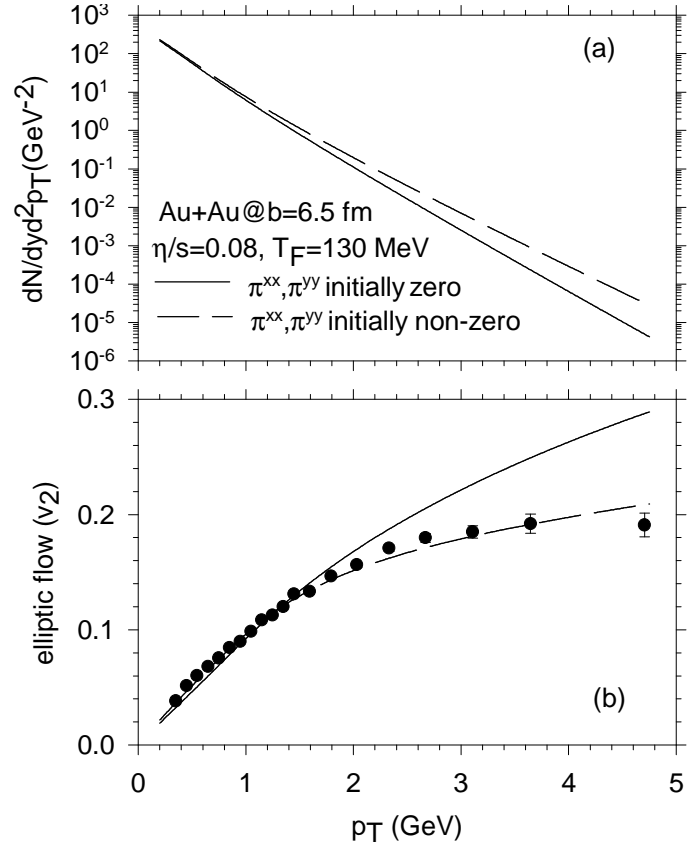}
\caption{ \label{F29} 
(a) Minimally viscous hydrodynamic predictions for the transverse momentum spectra of $\pi^-$, in  Au+Au collisions at impact parameter b=6.5 fm, for two values of initial shear stress tensors. The solid line corresponds to initial zero shear stress tensors $\pi^{xx}$=$\pi^{yy}$=$\pi^{xy}$=0. The dashed line corresponds to initial boost invariant values for the shear stress tensors, $\pi^{xx}$=$\pi^{yy}$=$2\eta/\tau_i$, $\pi^{xy}$=0.   Pion yield decrease if initially shear stress tensors are zero. (b) same as in (a) but for the elliptic flow for $\pi^-$. Elliptic flow increases if initially shear stress tensor is zero. The filled circles are the PHENIX data \cite{Adler:2004cj} on the transverse momentum dependence of elliptic flow in 13-26\% centrality Au+Au collisions.} 
\end{figure}

\begin{figure}[ht]
 \includegraphics[bb= 21 268 506 748
 ,width=0.5\linewidth,clip]{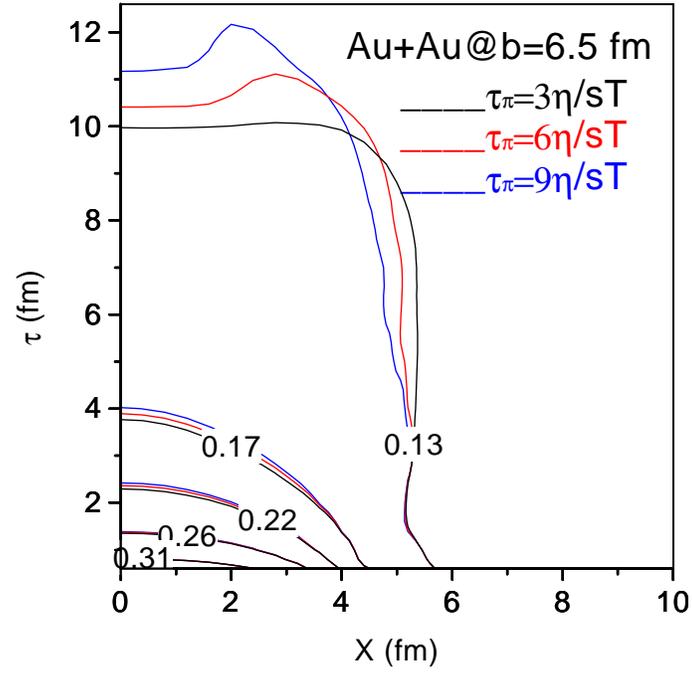}
\caption{\label{F30}(color online). Dependence of fluid evolution on the relaxation time $\tau_\pi$. The contours of constant temperature in   $x-\tau$ plane at a fixed y=0 fm, in Au+Au collision at b=6.5 fm are shown. The blue, red and black lines corresponds to  relaxation time $\tau_\pi$=$3\eta/sT$, $6\eta/sT$ and $9\eta/sT$. respectively.}
\end{figure}

\begin{figure}
\includegraphics[bb=35 189 533 809 
    ,width=0.5\linewidth,clip]{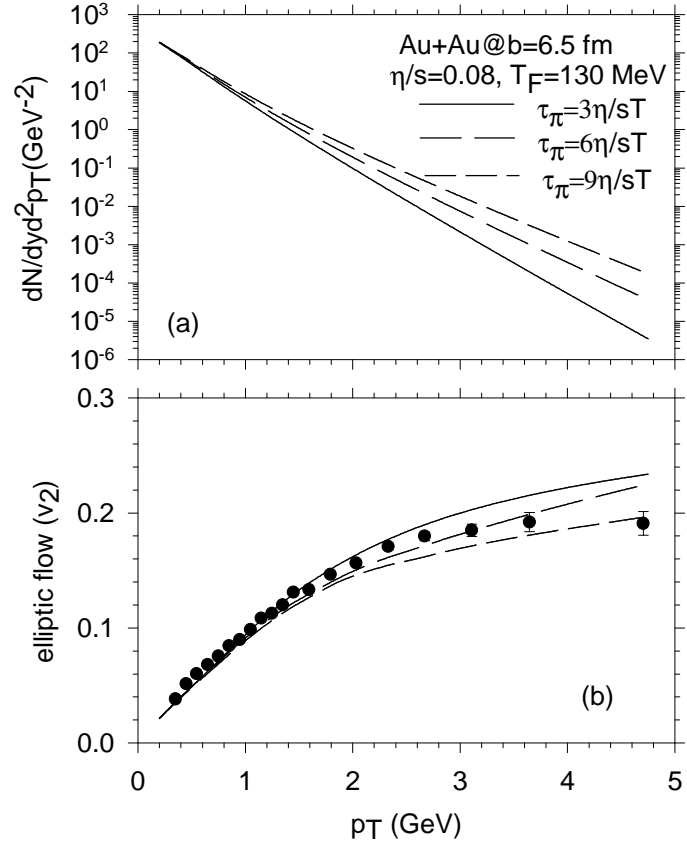}
\caption{\label{F31}
(a) Minimally viscous hydrodynamic predictions for the transverse momentum spectra of $\pi^-$, in  Au+Au collisions at impact parameter b=6.5 fm. The solid, dashed and short dashed lines corresponds to relaxation time, $\tau_\pi$=$3\eta/sT$, $6\eta/sT$ and $9\eta/sT$. Pion yield increase as the relaxation time increase. (b) same as in (a) but for the elliptic flow for $\pi^-$. Elliptic flow decreases with increasing relaxation time. The filled circles are the PHENIX data \cite{Adler:2004cj}  on the transverse momentum dependence of elliptic flow in 13-26\% centrality Au+Au collisions.}
\end{figure}

Fig.24:(color online). Same as in Fig.\ref{F23} but for $K^+$. 

Fig.25:(color online). Same as in Fig.\ref{F23} but for proton. 

Fig.26: PHENIX data on the centrality dependence of average $p_T$ for $\pi^-$, $K^+$ and protons are shown. The solid lines are predictions from minimally viscous dynamics. 

Fig.27:(color online). Evolution of shear stress tensor $\pi^{xx}(x,y=0)$ for x=0,1,3 and 6 fm. The blue and black lines corresponds to fluid evolution with zero and non-zero (boost-invariant value) initial shear stress tensors. 

Fig.28:(color online). Dependence of fluid evolution on initial shear stress tensor. Constant temperature contours in $x-\tau$ plane at a fixed y=0 fm, in a Au+Au collision at b=6.5 fm are shown. The black lines correspond to initially zero shear stress tensor. The blue lines correspond to initially non-zero,  boost-invariant shear stress tensors.    

Fig.29:(a) Minimally viscous hydrodynamic predictions for the transverse momentum spectra of $\pi^-$, in  Au+Au collisions at impact parameter b=6.5 fm, for two values of initial shear stress tensors. The solid line corresponds to initial zero shear stress tensors $\pi^{xx}$=$\pi^{yy}$=$\pi^{xy}$=0. The dashed line corresponds to initial boost invariant values for the shear stress tensors, $\pi^{xx}$=$\pi^{yy}$=$2\eta/\tau_i$, $\pi^{xy}$=0.   Pion yield decrease if initially shear stress tensors are zero. (b) same as in (a) but for the elliptic flow for $\pi^-$. Elliptic flow increases if initially shear stress tensor is zero. The filled circles are the PHENIX data \cite{Adler:2004cj} on the transverse momentum dependence of elliptic flow in 13-26\% centrality Au+Au collisions. 

Fig.30:(color online). Dependence of fluid evolution on the relaxation time $\tau_\pi$. The contours of constant temperature in   $x-\tau$ plane at a fixed y=0 fm, in Au+Au collision at b=6.5 fm are shown. The blue, red and black lines corresponds to  relaxation time $\tau_\pi$=$3\eta/sT$, $6\eta/sT$ and $9\eta/sT$. respectively. 

Fig.31:(a) Minimally viscous hydrodynamic predictions for the transverse momentum spectra of $\pi^-$, in  Au+Au collisions at impact parameter b=6.5 fm. The solid, dashed and short dashed lines corresponds to relaxation time, $\tau_\pi$=$3\eta/sT$, $6\eta/sT$ and $9\eta/sT$. Pion yield increase as the relaxation time increase. (b) same as in (a) but for the elliptic flow for $\pi^-$. Elliptic flow decreases with increasing relaxation time. The filled circles are the PHENIX data \cite{Adler:2004cj}  on the transverse momentum dependence of elliptic flow in 13-26\% centrality Au+Au collisions.

\eject


 \clearpage



 \clearpage

\clearpage

\end{document}